\documentclass[12pt]{article}

\usepackage{amsmath}
\usepackage{amssymb}

\newtheorem{theorem}{Theorem}[section]

\newtheorem{corollary}{Corollary}[section]
\newtheorem{prop}{Proposition}[section]
\newtheorem{definition}{Definition}[section]
\numberwithin{equation}{section}

\begin{document}

\begin{center}

On Segal-Bargmann analysis for finite Coxeter groups \\
and its heat kernel

\end{center}

\vskip .5cm

\centerline{Stephen Bruce Sontz\footnote{Research partially
supported by CONACYT (Mexico) project 49187.}}

\centerline{Centro de Investigaci\'on en Matem\'aticas, A.C. (CIMAT)}

\centerline{Guanajuato, Mexico}

\centerline{email: sontz@cimat.mx}

\vskip .5cm

\begin{abstract}
\noindent
We prove identities involving the integral kernels of three versions
(two being introduced here) of the Segal-Bargmann transform
associated to a finite Coxeter group acting on a finite dimensional,
real Euclidean space (the first version essentially having been
introduced around the same time
by Ben~Sa\"id and {\O}rsted and independently by Soltani) and the Dunkl heat
kernel, due to R\"osler, of the Dunkl Laplacian associated with the same Coxeter group.
All but one of our relations are originally due to
Hall in the context of standard Segal-Bargmann analysis
on Euclidean space.
Hall's results (trivial Dunkl structure and arbitrary finite dimension)
as well as our own results in
 $\mu$-deformed quantum mechanics (non-trivial Dunkl structure, dimension one)
are particular cases of the results proved here.
So we can understand all of
these versions of the Segal-Bargmann transform 
associated to a Coxeter group
as Hall type transforms.
In particular, we define an analogue of Hall's Version~C generalized
Segal-Bargmann transform which is then shown to be Dunkl convolution with the Dunkl heat kernel
followed by analytic continuation.
In the context of Version~C we also introduce a new Segal-Bargmann space
and a new transform associated to the Dunkl theory.
Also we have what appears to be a new relation in this context between the Segal-Bargmann kernels
for Versions A and C.
\end{abstract}



\section{Introduction}
\label{sec1}

Since the introduction by Bargmann in \cite{BA} and Segal in \cite{SEG} in the early
1960's of a certain Hilbert space of holomorphic functions and an associated integral
kernel transform,
there has been much research on various deformations and generalizations
of what could be called Segal-Bargmann analysis, namely, the study of
this space and its transform.
These generalized Segal-Bargmann spaces and transforms are often quite
algebraic in nature.
(See the texts by Ali, et~al. \cite{ALI} and  by Perelomov \cite{PE}
and references therein for more details. Note that the transform is often called
a {\em coherent state transform}.)
However, Hall in \cite{HA} introduced a Segal-Bargmann space and transform
for compact Lie groups (and other closely related differential manifolds)
that has an analytic flavor, since it is based directly on the heat kernel
analysis associated with naturally given Laplacian operators.

   Meanwhile, in a development of modern harmonic analysis  starting in the late 1980's,
Dunkl introduced deformations of the partial derivatives and of the Fourier
transform in $\mathbb{R}^N$,
based on a finite Coxeter group generated by reflections in $\mathbb{R}^N$.
References to the original articles of Dunkl and more recent work are presented
in \cite{MR}.
However, this research was mainly focused on the configuration space $\mathbb{R}^N$,
while Segal-Bargmann analysis also involves the phase space, which in this case
is $\mathbb{C}^N$.
More recently Soltani in \cite{SO} and independently Ben~Sa\"id and {\O}rsted in \cite{SBSBO}
have introduced a Hilbert space of holomorphic functions on $\mathbb{C}^N$ and an associated
Segal-Bargmann type transform in the context of the Dunkl theory on $\mathbb{R}^N$.
(Be warned that what all of these authors call a ``generalized Fock space'' is for us
a generalized Segal-Bargmann space.
Also the ``chaotic transform'' of Soltani in \cite{SO} is a generalized Segal-Bargmann
transform for us.)
This generalizes a setup for dimension $N=1$ studied by Rosenblum in \cite{RO} and
by his student Marron in \cite{MA}.
The author and various collaborators have also worked extensively in recent years on this formulation
in dimension one and continue to do so.
(See \cite{SBS} and references therein.)
We refer to this special one-dimensional case of Dunkl theory as $\mu$-deformed quantum mechanics,
since it originally appeared in a paper of Wigner \cite{WI} concerning a question in the theory
of quantum mechanics.

     In this article we develop further the Segal-Bargmann analysis associated to the
Dunkl theory in $\mathbb{R}^N$ for any finite integer $N \ge 1$.
We prove various relations with the Dunkl heat kernel
for three \textit{versions} of this theory  in
Theorems \ref{version_a}, \ref{version_b} and \ref{version_c}.
These are called Versions A, B and C.
Version~A was essentially introduced in \cite{SBSBO}
and independently in \cite{SO} while the other two versions are being introduced here.
The idea of versions (or perhaps a better word, such as \textit{variants}) 
of Segal-Bargmann analysis was first introduced by Hall in \cite{HA} in the context
of his work on Segal-Bargmann analysis for compact Lie groups and other related
differential manifolds.

Most of our results were first
proved in arbitrary dimension $N \ge 1$ by Hall in \cite{HA} but with the trivial Dunkl structure,
and more recently we have shown these results in \cite {SBS} in the case of
non-trivial Dunkl structure but in dimension one.
So the present work generalizes both Hall's work and ours.
In particular, the generalized Segal-Bargmann transforms studied here are Hall type transforms.
We should note that a Version D was introduced in \cite{SBS}, and this can
also be done here in the same way, namely, by a change of measure unitary transformation
starting from Version C.
However, we do not include this minor detail here.
More importantly,
in the context of Version~C we define a new Segal-Bargmann space and a new transform
associated to the Dunkl theory.
This Version~C generalized Segal-Bargmann transform turns out to be equal to
a Dunkl convolution with a Dunkl heat kernel followed by analytic continuation.
We also find what seems to be a new relation (see equation (\ref{C-A2}))
that holds for any Dunkl structure and any dimension $N \ge 1$.

The ``correct'' manner of introducing holomorphic spaces into the Dunkl theory, in particular
the appropriate Segal-Bargmann space and its associated integral transform, is by no means obvious.
The approach in \cite{SBSBO} and \cite{SO} is to use the Dunkl kernel function to define a
reproducing kernel function which in turn defines a Hilbert space of holomorphic functions.
(See Theorem \ref{thm1}.)
As is well known, this gives a unique space.
Then one could deal with the problem of defining a unique Segal-Bargmann transform.
However, we want to define three different versions of the Segal-Bargmann transform.
So we would like to find a unifying concept that uniquely determines each of those three transforms.
Now the combined results of Theorems \ref{version_a}, \ref{version_b} and \ref{version_c}
do show that there is a unifying global structure, namely the Dunkl heat kernel.
This says that the ``correct'' transform is found in \cite{SBSBO} and \cite{SO} and
that it corresponds to our Version~A.
(See Corollary~\ref{big_cor}.)
The importance of the heat kernel in Segal-Bargmann analysis,
originally due to Hall in \cite{HA}, is a leitmotif of this article.
Even though this global structure, the Dunkl heat kernel,
was identified in the case $N=1$ (with Coxeter
group $\mathbb{Z}_2$) in \cite{SBS}, it is still surprising to find these same relations
again in the very general case of any finite dimension and any Coxeter group.
Moreover, as we have already noted, this also leads in a natural way to a new Segal-Bargmann
space and associated transform for Version~C.
(See Theorem \ref{version_c}.)

     Although we will not use this here, we would like to note that
the Dunkl theory has close connections with probability theory as
first developed in \cite{MV} by R\"osler and Voit.
They show that the Dunkl Laplacian is the generator of a strongly continuous Markov semigroup
and study its associated stochastic process (a generalized
Brownian motion with jump discontinuities).
More recent work by Gallardo and Yor on this subject is in \cite{GY1} and \cite{GY2}.
Also there is some very recent related work in \cite{DEM1} and \cite{DEM2} by Demni.

Another relation of this material to classical mathematics can be seen in the case
of dimension $N=1$.
In that case the Dunkl Laplacian
(see Def.~\ref{dunkl_lap_op} here or Eq.~(2.5.1) in \cite{RO}) for even functions
$f : \mathbb{R} \to \mathbb{C}$
is given by $f^{\prime\prime}(q) + (2 \mu /q) f^{\prime}(q)$,
where $\mu > -1/2$ and $q \in \mathbb{R}$.
This should be compared with the radial part of the usual Laplacian in $\mathbb{R}^d$,
\begin{equation}
\label{euclid_lap_rad}
        \dfrac{\partial^2}{\partial r^2} + \dfrac{d-1}{r} \dfrac{\partial}{\partial r},
\end{equation}
where $r = ||x||$ for $x \in \mathbb{R}^d$ is the usual radial variable.
Thus when $2\mu +1$ is an integer $d \ge 1$ (equivalently, $\mu \in \{ 0, 1/2, 1 , 3/2, \dots \}$,
i.e., $\mu$ is a half integer),
we can identify the Dunkl Laplacian
with the radial part of the Euclidean Laplacian, while for other values of $\mu > -1/2$
we can think of the Dunkl Laplacian (at least in its action on even functions) as a continuous family
of operators that interpolates the discrete family (\ref{euclid_lap_rad}).
Of course, even functions
$f : \mathbb{R} \to \mathbb{C}$ are in one-to-one and onto correspondence with the functions
$g :[0, \infty) \to \mathbb{C}$ by the restriction $ f \mapsto g := f \upharpoonright_{[0,\infty)}$.
The radial functions on $\mathbb{R}^d$ are also clearly in one-to-one and onto correspondence with the
functions $g :[0, \infty) \to \mathbb{C}$.
Using these identifications, the  Dunkl Laplacian on even function for half integers $\mu$
corresponds to the Euclidean Laplacian on radial functions.
Since the Segal-Bargmann theory is built on the heat kernel which in turn comes from
the appropriate Laplacian, we see that the Segal-Bargmann theory of
a certain class of functions (namely, the even functions or, equivalently,
the $\mathbb{Z}_2$-invariant functions) in dimension one with half integers $\mu$ 
is identified with the usual Segal-Bargmann theory in a Euclidean space of a certain
class of functions (namely, the radial ones).

Something similar also happens in higher dimension.
Consider the setup considered in \cite{BSO_FLAT}, where $G$ is a semisimple, connected
Lie group with finite center and $K$ is a maximal compact subgroup of $G$.
Then the quotient space $G/K$ is a Riemannian symmetric space of non-compact type.
Then the radial part of the Laplace-Beltrami on $G/K$
is given in the notation of \cite{BSO_FLAT}  by
\begin{equation}
\label{LsubA}
      \mathcal{L}_A + \sum_{\alpha \in \Sigma^+} m_\alpha (\coth \alpha) A_\alpha
\end{equation}
where $\mathcal{L}_A $ is the Laplace-Beltrami operator on $A$,
the connected subgroup of $G$ associated to a maximal abelian subspace
$\mathcal{A}$ of $\mathcal{P}$.
Here $ \mathcal{G} = \mathcal{K} \oplus \mathcal{P}$ is the Cartan decomposition
of the Lie algebra $\mathcal{G}$ of $G$.
(For the rest of this notation and further discussion, see \cite{BSO_FLAT} and \cite{HE}.)
This operator also is given in Equation (1.3) and analyzed in~Example~1.5 in \cite{OS}.
The point is that this formula compares favorably with
the Dunkl Laplacian $\Delta_{\mu}$ (see Definition \ref{dunkl_lap_op})
when applied to a Coxeter group invariant function $f$, which in our notation is
\begin{equation}
\label{delta_mu_inv}
 (\Delta_{\mu} f)(x) = \Delta f (x) + 2
 \sum_{\alpha \in \mathcal{R}^+} \mu (\alpha)
\dfrac{\left\langle \alpha , \mathrm{grad} f(x) \right\rangle}{\left\langle \alpha , x \right\rangle}.
\end{equation}
(See the formula on p.~286 of \cite{SBSBO}.)
After decoding all the notation in these two formulas, the only real difference is that the
former has a factor of $\coth \alpha $ where the later has a factor of $1/\alpha$, taking
the root $\alpha$ to be an element in the appropriate \textit{dual} vector space,
which will not be our convention in the rest of this article.
Now $1/z$ is the principal part of the odd meromorphic function $\coth z$
and so can be considered as the approximation of $\coth z$ for $z \in \mathbb{C}$ in an
infinitesimal neighborhood of $0$.
So we can think of (\ref{delta_mu_inv}) as an infinitesimal approximation of (\ref{LsubA})
provided that we take $x$ in (\ref{delta_mu_inv}) to be an element in the Euclidean space
$\mathcal{A}$.
This provides another link between the theory presented here and classical analysis.
I thank B.~Hall for bringing the material in this and the previous paragraph to my attention.

The article is organized as follows.
Section~\ref{sec2} has a detailed exposition of known results,
which form the background material for the rest of the article.
We establish there our own
notation and conventions and, in particular, the use (due to Hall in \cite{HA})
of the real parameter $t>0$
for the time parameter of the heat equation (and its associated kernel)
as well as for the quantum deformation parameter, namely, Planck's constant.
In Section~\ref{sec3}
we show how this parameter is related to the dilation structures of the
configuration space $\mathbb{R}^N$  and of the phase space $\mathbb{C}^N$.
Then our results concerning the relations between the three
kernels of the Segal-Bargmann transforms associated to a Coxeter group
and the Dunkl heat kernel are presented and proved.
We conclude with some comments, mostly having to do
with possible avenues for further research.

\section{Background material}
\label{sec2}

We discuss here the preliminaries needed for the proofs
in Section~\ref{sec3} of the results described in the abstract.
We do not present in this section  all of the proofs, since these are  all known results.
References for this background material are \cite{SBSBO},
\cite{HU}, \cite{MR}, \cite{MR2}, \cite{SO} and \cite{TX}.
These may be consulted for more details and proofs.
We warn the reader that some of our notation and normalizations are not standard.

We let $\mathbb{R}^N$ denote the Euclidean space of finite dimension $N \ge 1$
with its standard inner product
$$
\left\langle x,y \right\rangle = \sum_{j=1}^N x_j y_j
$$
for $x = (x_1, \dots , x_N), y = (y_1, \dots , y_N) \in \mathbb{R}^N$.
Also we let $ || x || = \left\langle x,x \right\rangle^{1/2}$ denote
the standard Euclidean norm and let $x^2 = \langle x,x \rangle$.

For any $0 \ne \alpha \in  \mathbb{R}^N$, we denote by
$\sigma_\alpha$ the orthogonal reflection in the hyperplane
orthogonal to $\alpha$, that is
$\{ \alpha \}^\perp$.
Explicitly, we have the formula
$$
 \sigma_\alpha(x) = x -
\frac{2\left\langle \alpha,x \right\rangle}{|| \alpha ||^2} \alpha
$$
for all $x \in \mathbb{R}^N$.
One easily shows that
$\langle \sigma_\alpha(x), \sigma_\alpha(y) \rangle = \langle x,y \rangle$
holds for all $x, y \in \mathbb{R}^N$, that is,
$\sigma_\alpha \in O (\mathbb{R}^N)$, the orthogonal group of
$\mathbb{R}^N$.
Clearly, among other elementary properties, we have that
$\det \, \sigma_\alpha = -1$ and $\sigma_\alpha^2 = I$, the identity.

\begin{definition}
A finite set $\mathcal{R}$ of non-zero vectors in the Euclidean space $\mathbb{R}^N$
is called a {\em root system} if
\begin{enumerate}

\item $\alpha \in \mathcal{R} \Rightarrow -\alpha \in \mathcal{R}$,

\item $\alpha \in \mathcal{R}$ and $r \alpha \in \mathcal{R}$ for
some $r \in \mathbb{R} \Rightarrow r = \pm 1$,

\item $\sigma_\alpha(\mathcal{R}) = \mathcal{R}$
for all $\alpha \in \mathcal{R}$.

\end{enumerate}
(While Property~3 here implies Property~1, a little redundancy does no harm.)

We also follow a standard convention by requiring that each root $ \alpha \in \mathcal{R}$
be normalized by $|| \alpha ||^2 =2 $.
This has no real effect on the results while it allows for some degree of compatibility with
other authors and simplifies some formulas.

Given such a root system $\mathcal{R}$, we denote the subgroup
of $O(\mathbb{R}^N)$ generated
by the elements $\sigma_\alpha$ for all $\alpha \in \mathcal{R}$
as $ G \equiv G(\mathcal{R})$.
(It turns out that this group is finite as we shall soon see.)
We say that $G$ is the
{\em (finite) Coxeter group} associated with the root system $\mathcal{R}$.

A $G$-invariant function $\mu : \mathcal{R} \to \mathbb{C}$ is
called a {\em multiplicity function}.
(Note that we are using Property~3 above of a root system here, since we
are requiring that $\mu (g \alpha) = \mu (\alpha)$ for $g\in G$ and
$\alpha \in \mathcal{R}$ and so we need to know that $g \alpha \in \mathcal{R}$.)

\end{definition}

To see that $G$ is finite, we note that each element $g \in G$ acts on the finite
set $\mathcal{R}$, and so we have a homomorphic image of $G$ realized
as a subgroup of the finite group of permutations of $\mathcal{R}$.
If we can show that this homomorphism is injective, then
we will have that $G$ is finite.
But if $g\in G$ acts as the identity on $\mathcal{R}$, then
it acts as the identity on the subspace generated by $\mathcal{R}$,
namely $ \mathrm{span} (\mathcal{R})$.
But {\em every} element of $G$ acts as the identity on the
orthogonal complement of $ \mathrm{span} (\mathcal{R})$.
So $g$ acts as the identity on all of $\mathbb{R}^N$, that is,
it is the identity element of $O (\mathbb{R}^N)$ and
so also the identity element of the subgroup $G$.
And this proves that the kernel of the homomorphism is
trivial, and so we have the desired injectivity.

We note that according to our definition $\mathcal{R} = \emptyset $, the empty set, is a root system
whose associated group consists of exactly one element (the identity), and
therefore the trivial subgroup of $O (\mathbb{R}^N)$ is a Coxeter group.
Non-trivial examples of root systems and their associated Coxeter groups
are given in \cite{MR}.
Also the text \cite{HU} has a wealth of information on Coxeter groups.

We let $\mathcal{R}^+$ denote any subset of positive elements
in $\mathcal{R}$ with respect to a given total order on $\mathbb{R}^N$.
(An {\em order} on $\mathbb{R}^N$ is a partial order $<$ such that
$ u < v \Rightarrow u +w < v + w$ and $ru < rv$ for all
$u,v,w \in \mathbb{R}^N$ and $r>0$.
Such an order is said to be {\em total} if for all $u,v \in \mathbb{R}^N$
either $u < v$ or $ v < u$ or $ u = v$.)
We define various objects in terms of the subset $\mathcal{R}^+$ of positive
elements with respect to a given total order (which {\em do} exist),
since this is how it is usually done in the literature.
However, none of these depends on the particular choice of total order.
This is due to various basic facts the reader can verify such as
$\mathcal{R}  = \mathcal{R}^+ \cup (-\mathcal{R}^+)$
(a disjoint union),  $\mathbb{R} \alpha \cap \mathcal{R} = \{ \alpha, -\alpha\} $
for all $\alpha \in \mathcal{R}$, $\sigma_{-\alpha} = \sigma_{\alpha}$,
$\sigma_\alpha(\alpha) = - \alpha$ and
$\mu(-\alpha) = \mu(\alpha)$.

\begin{definition}
For any $\xi \in \mathbb{R}^N$ and multiplicity function
$\mu : \mathcal{R} \to \mathbb{C}$ we define the {\em Dunkl operator} $ T_{\xi,\mu}$ by
$$
  T_{\xi,\mu}f(x):= \partial_\xi f(x) +
  \sum_{\alpha \in \mathcal{R}^+} \mu(\alpha) 
  \frac{\langle \alpha, \xi \rangle}{\langle \alpha , x \rangle} \cdot
\big( f(x) - f(\sigma_\alpha(x)) \big),
$$
where $\partial_\xi = \langle \xi , {\rm grad} \rangle$ is the directional derivative
associated to $\xi \in \mathbb{R}^N$
(with ${\rm grad} = (\partial/\partial x_1, \dots , \partial/\partial x_N)$
being the usual gradient operator),
 $ x \in \mathbb{R}^N$
and $f \in C^1 (\mathbb{R}^N)$ is a complex valued function.
This definition can equivalently be written as
$$
  T_{\xi,\mu}f(x)= \partial_\xi f(x) + \dfrac{1}{2}
  \sum_{\alpha \in \mathcal{R}} \mu(\alpha)
  \frac{\langle \alpha, \xi \rangle}{\langle \alpha , x \rangle} \cdot
\big( f(x) - f(\sigma_\alpha(x)) \big),
$$
which shows that this operator does not depend on the choice of the total order.
\end{definition}

Note that the linear operator $T_{\xi,\mu}$ depends linearly on $\xi$.
A non-trivial result here is that the operators $T_{\xi,\mu}$ and $T_{\eta, \mu}$ commute
for all $\xi, \eta \in \mathbb{R}^N$.
For the constant multiplicity function $\mu \equiv  0$
or for $\mathcal{R} = \emptyset$ (either case being called the \textit{trivial Dunkl structure})
the operator $T_{\xi,\mu}$ reduces to the directional derivative $\partial_\xi$
associated to the vector $\xi \in \mathbb{R}^N$.

\begin{definition}
\label{dunkl_lap_op}
Suppose that $\xi_1, \dots , \xi_N$ is any orthonormal basis of $\mathbb{R}^N$
and that $\mu : \mathcal{R} \to \mathbb{C}$ is any multiplicity function.
Then we define the {\em Dunkl Laplacian} by
$$
             \Delta_\mu : = \sum_{j=1}^N (T_{\xi_j,\mu})^2,
$$
say as an operator acting on the domain $C^2(\mathbb{R}^N)$.
(It turns out that $\Delta_\mu$ does not depend on the choice of
orthonormal basis.)

\end{definition}

For the rest of this article we will assume that the multiplicity
function is non-negative: $\mu \ge 0$.
Now we consider the problem of solving
\begin{gather*}
T^x_{\xi,\mu}E_\mu(x,y) = \langle \xi, y \rangle E_\mu(x,y) \\
E_\mu(0,y) = 1
\end{gather*}
for all $x,y, \xi \in \mathbb{R}^N$.
Here the superscript $x$ in $T^x_{\xi,\mu}$ indicates that the
operator acts on the first variable of $E_\mu$.

This problem has a unique real analytic solution
$E_\mu : \mathbb{R}^N \times \mathbb{R}^N \to \mathbb{R}$,
which extends to a holomorphic function
$E_\mu : \mathbb{C}^N \times \mathbb{C}^N \to \mathbb{C}$.
We call this function $E_\mu$ the {\em Dunkl kernel function}.
If $\mu \equiv 0$, then
$\Delta_\mu = \Delta_0 = \sum_{j=1}^N {\partial^2}/{\partial x_j^2}$
is the usual Euclidean Laplacian.
So we have $E_0(x,y) = e^{\langle x,y \rangle}$ for
$x,y \in \mathbb{R}^N$ and
$E_0(z,w) = e^{\langle z,w \rangle}$ for
$z,w \in \mathbb{C}^N$, where $\langle z,w \rangle$ is \textit{bilinear}
over the complex field $\mathbb{C}$ and extends the Euclidean 
inner product $\langle x,y \rangle$.
Some of the properties of the Dunkl kernel are
\begin{gather*}
E_\mu(z,w) = E_\mu(w,z), \\
E_\mu(gx, gw) = E_\mu (z,w), \\
E_\mu(\lambda z, w ) = E_\mu ( z, \lambda w),
\end{gather*}
where $z,w \in \mathbb{C}^N$, $g \in G$ and $\lambda \in \mathbb{C}$.

   The Segal-Bargmann space associated to a finite Coxeter group
is defined in \cite{SBSBO} for a specific value of Planck's constant.
We present their results in the next theorem, but with an arbitrary
value $t>0$ of Planck's constant.

\begin{theorem} 
\label{thm1}
(Ben~Sa\"id, {\O}rsted \cite{SBSBO})
Suppose that the multiplicity function satisfies $\mu \ge 0$ and
that $t>0$ is given.
Define $\mathbb{K}_{\mu,t} : \mathbb{C}^N \times \mathbb{C}^N \to \mathbb{C}$ by
$$
\mathbb{K}_{\mu,t}(z,w):=E_{\mu} \left( \dfrac{z^{*}}{ t^{1/2} },\dfrac{w}{ t^{1/2} } \right)
$$
for $z,w\in \mathbb{C}^N$ and $z^{*}:= (z_1^*, \dots , z_N^*)$, where $z= (z_1, \dots , z_N)$
with each $z_j \in \mathbb{C}$ and $\lambda^*$ is the complex conjugate of $\lambda \in \mathbb{C}$.
\begin{enumerate}
 \item $\mathbb{K}_{\mu,t}$ is positive definite, that is,
\begin{equation*}
\sum_{i,j=1}^m a_i^* \mathbb{K}_{\mu,t} ( z^{(i)},  z^{(j)} ) a_j \ge 0
\end{equation*}
for every integer $m \ge 1$ and $a_1, \dots , a_m \in \mathbb{C}$
and $ z^{(1)}, \dots,  z^{(m)} \in \mathbb{C}^N$.
 \item There exists a reproducing kernel Hilbert space $\mathcal{B}_{\mu,t}$ of
holomorphic functions $f: \mathbb{C}^N \to \mathbb{C}$ whose
reproducing kernel function is $\mathbb{K}_{\mu,t}$.
\item The set $\mathcal{P} (\mathbb{C}^N)$ of all holomorphic polynomials
in $z_1, \dots , z_N$ is a dense subspace of $\mathcal{B}_{\mu,t}$.
\end{enumerate}
\end{theorem}

We say that $\mathcal{B}_{\mu,t}$ is {\em the Segal-Bargmann space associated
to the finite Coxeter group} $G(\mathcal{R}) $
generated by the root system $\mathcal{R}$
and the non-negative multiplicity function $\mu : \mathcal{R} \to [0, \infty)$
for the value $t>0$ of Planck's constant.
Though we should include $\mathcal{R}$ in the notation for this Hilbert space, we omit it since
we always have one fixed root system $\mathcal{R}$ in our discussions.

Let us note that in \cite{SBSBO} the proofs of these statements are given for
the particular case $t=1$.
To extend these proofs to the general case $t>0$ is straightforward.
The space $\mathcal{B}_{\mu,t}$ for $t=1$  was also introduced
independently by Soltani in \cite{SO}.

In the one dimensional case ($N=1$) all of the non-empty root systems have the form
$\mathcal{R}= \{\alpha, -\alpha \}$ for some real number $\alpha >0$
(dropping the normalization condition $|| \alpha || = 2 $ for this example).
The associated Coxeter group is generated by $\sigma_\alpha = \sigma_{-\alpha}$,
which is the only non-trivial element in
$O (\mathbb{R}) \cong \mathbb{Z}_2$,
namely, the reflexion in the origin $ x \mapsto -x$ for $x \in \mathbb{R}$.
So the finite Coxeter group is $\mathbb{Z}_2$, the cyclic group of order $2$, in this case.
In \cite{SBSBO} the authors identify the associated Segal-Bargmann
space in Example~4.17.
This turns out to be exactly the $\mu$-deformed Segal-Bargmann
space studied by us in \cite{SBS}.
See \cite{SBSBO} and \cite{SBS} and references therein for more details.
Notice that for dimension $N > 1 $ the orthogonal group $O (\mathbb{R}^N)$
is infinite, and thus in that case no Coxeter group can be equal to $O (\mathbb{R}^N)$.
So the case $N=1$ is exceptional in this regard.

We denote the inner product on $\mathcal{B}_{\mu,t}$
by $\langle\langle \cdot \, , \cdot \rangle\rangle_{\mu,t}$.
In particular, if we define
$$
   \mathbb{K}_{z,\mu,t}(w):=  \mathbb{K}_{\mu,t}(z,w)
$$
for $z,w \in \mathbb{C}^N$,
then we have that $\mathbb{K}_{z,\mu,t} \in \mathcal{B}_{\mu,t}$ and that
\begin{equation*}
\langle\langle  \mathbb{K}_{z,\mu,t} \, , f \rangle\rangle_{\mu,t} = f(z)
\end{equation*}
for all $f \in \mathcal{B}_{\mu,t}$ and $z \in \mathbb{C}^N$,
which is the reproducing property of the kernel function.
Note that we follow the physics convention that an inner product is anti-linear
in its first argument and linear in its second.

Having defined $\mathcal{B}_{\mu,t}$ as a Hilbert space of
certain holomorphic functions
on the \textit{phase space} $\mathbb{C}^N$, we would now like to define a Hilbert space of functions
on the \textit{configuration space} $\mathbb{R}^N$ and define an associated Segal-Bargmann
type transform from the latter Hilbert space to the former.

First, we define a weight function $\tilde{\omega}_{\mu,t} : \mathbb{R}^N \to [0,\infty)$ by
$$
  \tilde{\omega}_{\mu,t} (x):= t^{-(\gamma_\mu + N/2)} \prod_{\alpha \in \mathcal{R}^+}
  | \langle \alpha , x \rangle |^{2\mu(\alpha)} =
  t^{-(\gamma_\mu + N/2)} \prod_{\alpha \in \mathcal{R}}
  | \langle \alpha , x \rangle |^{\mu(\alpha)}
$$
for $x \in \mathbb{R}^N$, where $t>0$, $\, \mu \ge 0$ is a multiplicity function and
$$\gamma_\mu:= \sum_{\alpha \in \mathcal{R}^+} \mu(\alpha)
= \dfrac{1}{2} \sum_{\alpha \in \mathcal{R}} \mu(\alpha).
$$
If we put $t=1$, the first expression for $\tilde{\omega}_{\mu,t}$ is the one usually given.
The second expression shows that this weight function does not depend on
the particular choice of the total order.
Similarly, the first expression for $\gamma_\mu$ is the usual one, while the second shows that
this quantity does not depend on the particular choice of the total order.

We use this weight to define the measure
$
 \mathrm{d} \, \tilde{\omega}_{\mu,t}(x) := \tilde{\omega}_{\mu,t}(x) \, \mathrm{d}^N x
$
on $\mathbb{R}^N$, where $\mathrm{d}^N  x$ denotes Lebesgue measure on $\mathbb{R}^N$.
We also define the dilated {\em Macdonald-Mehta-Selberg constant} for $t > 0$ by
$$
 c_{\mu,t} := \int_{\mathbb{R}^N} \mathrm{d} \, \tilde{\omega}_{\mu,t} (x) \, e^{-x^2/2t}.
$$
This is usually defined only for $t=1$.
However, $c_{\mu,t} = c_{\mu,1} \equiv c_\mu$ so that the dilation parameter is unimportant.
Clearly we have $0 < c_\mu < \infty $.

Next we define the weight function that we will mainly use from now on:
$$
     \omega_{\mu,t} (x):= c_\mu^{-1} \tilde{\omega}_{\mu,t}(x) = c_\mu^{-1}
  t^{-(\gamma_\mu + N/2)} \prod_{\alpha \in \mathcal{R}}
  | \langle \alpha , x \rangle |^{\mu(\alpha)}
$$
for $x \in \mathbb{R}^N$.
So we are introducing a normalization constant, namely $ c_\mu^{-1}$, that
is not always used by other authors, though it is used in \cite{SO}.
We feel that one advantage of this convention is that it 
puts the Macdonald-Mehta-Selberg constant in one place, rather than having it
appear in a large number of formulas.
As above, we use this weight function to define a measure on $\mathbb{R}^N$, namely
$
 \mathrm{d} \, \omega_{\mu,t}(x) := \omega_{\mu,t}(x) \, \mathrm{d}^N x.
$
Some basic properties of this weight function are
\begin{gather*}
\omega_{\mu,t} (gx)= \omega_{\mu,t} (x), \\
\omega_{\mu,t} (\lambda x) = |\lambda|^{2\gamma_{\mu} } \omega_{\mu,t} (x),
\end{gather*}
where $x \in \mathbb{R}^N$, $g \in G(\mathcal{R})$ and $\lambda \in \mathbb{R}$.

Then the Hilbert space of functions on the configuration space
$\mathbb{R}^N$ is simply defined as an $L^2$ space:
$$
 L^2 ( \mathbb{R}^N , \omega_{\mu,t}(x) \, \mathrm{d}^N x) \equiv
L^2 ( \mathbb{R}^N , \omega_{\mu,t} ) \equiv L^2 (\omega_{\mu,t}).
$$

We also have a sort of normalization condition for $\mathrm{d} \, \omega_{\mu,t}$, namely
\begin{equation}
\label{sono}
\int_{\mathbb{R}^N} \mathrm{d} \, \omega_{\mu,t} (x) \, e^{-x^2/2t} =1.
\end{equation}

We comment that $\{ \mathrm{d} \omega_{\mu,t} \}_{t \ge 0}$ is a natural family
of measures in this theory for at least three reasons.
First, the Dunkl operators are anti-symmetric in the $L^2$ spaces associated to each of these measures.
Second, each of these measures is a Haar type measure, that is, invariant under
the Dunkl translation operators (defined below).
The proof of the second assertion is similar to the proof of the special case of
it that is proved in \cite{SBS}.
Thirdly, this is the self-dual measure for the generalized Fourier
(or Dunkl) transform, defined later.

We remark that this Hilbert space does depend on the parameter $t > 0$ because of
the normalization factor in the weight $\omega_{\mu,t}$.
This is not the usual convention.
However, these configuration spaces are all equivalent.
We will use the following definition to show this.

\begin{definition}
Suppose that $\lambda >0 $, $x \in \mathbb{R}^N$ and $\psi : \mathbb{R}^N \to \mathbb{C}$.
Define the {\em dilation operator} by $\lambda$ as
$$
   \delta_\lambda \psi (x) := \psi ( \lambda x).
$$
Similarly, we define the {\em dilation operator} by
$\lambda \in \mathbb{C}^N \setminus \{ 0 \} $ as
$$
D_\lambda f(z) := f (\lambda z),
$$
where $z \in \mathbb{C}^N$ and $f: \mathbb{C}^N \to \mathbb{C}$.
However, we will only use this second definition for $\lambda > 0$.
\end{definition}

Then a change of variables shows that
$\delta_\lambda : L^2( \omega_{\mu,t}) \to L^2( \omega_{\mu,s})$
is a unitary map, provided that $ \lambda = (t/s)^{1/2}$.
Its inverse is clearly $\delta_{\lambda^{-1}}$.
The advantage of using the family of scaled measures $ \omega_{\mu,t}$
and spaces $ L^2(\omega_{\mu,t})$, where
$t > 0$, is that there is no Jacobian factor in the dilation map $\delta_\lambda$.
This makes for a close analogy with the dilations $D_\lambda$ on spaces of complex functions,
such as $\mathcal{B}_{\mu,t}$, where we do not necessarily have a measure nor a Jacobian.
We will develop this further in the next section in Proposition \ref{p1}.

We will consider a variation of the Hilbert space for the configuration space $\mathbb{R}^N$.
This is the {\em ground state} Hilbert space
$L^2(\mathbb{R}^N, m_{\mu,t})$, where we use the measure
$$
\mathrm{d} m_{\mu,t} (x) := e ^{-x^2/2t} \mathrm{d} \omega_{\mu,t} (x)
$$
for $x \in \mathbb{R}^N$, which is clearly a probability measure by (\ref{sono}).
By a change of variables argument, we see that
$\delta_{(t/s)^{1/2}} : L^2(\mathbb{R}^N, m_{\mu,t}) \to L^2(\mathbb{R}^N, m_{\mu,s})$
is a unitary isomorphism.

We define an inner product on the vector space over the reals,
$\mathcal{P}({\mathbb{R}^N})$, of
real valued polynomials in the variables $x_1, \dots , x_N$ by
\begin{gather*}
    [p,q]_{\mu,t} := \int_{\mathbb{R}^N}  \mathrm{d} m_{\mu,t} (x) \,
e^{-t\Delta_\mu/2} p (x) \, e^{-t\Delta_\mu/2} q (x)\\
 = \langle e^{-t\Delta_\mu/2} p, e^{-t\Delta_\mu/2} q \rangle_{L^2(m_{\mu,t})}
\end{gather*}
for $p,q \in \mathcal{P}({\mathbb{R}^N})$.
Since $\Delta_\mu$ maps a polynomial $p$ of degree $d \ge 2$ to
a polynomial of degree $d-2$ and annihilates all other polynomials, the infinite series
$$
  e^{-t\Delta_\mu/2} p = \sum_{k=0}^\infty \dfrac{1}{k!} \left(-\dfrac{t}{2}\right)^k \Delta_\mu^k p
$$
has only finitely many non-zero terms and so is also a polynomial of degree $d$.
An alternative expression for this inner product is the Fischer type formula
$$
[p,q]_{\mu,t} = \big( (\delta_{t^{1/2} } p) (T_\mu) (\delta_{t^{1/2} } q) \big)(0).
$$
Here $r(T_\mu)$ is defined for any polynomial $r \in \mathcal{P}({\mathbb{R}^N})$
to be the operator obtained by replacing each occurrence of $x_i$
in $r(x) = r(x_1, \dots, x_N)$ with the Dunkl operator $T_{\epsilon_i,\mu}$,
where $\{ {\epsilon_i}  \}_i$ is the standard basis of $\mathbb{R}^N$.
Then $r(T_\mu)$ is well defined since the $T_{\epsilon_i,\mu}$ form a commutative
family of operators.

For any $p \in \mathcal{P}({\mathbb{R}^N})$ we let $p$
also denote its unique analytic continuation to $\mathbb{C}^N$, that is to say, the
``same'' polynomial considered now as a holomorphic function of $z \in \mathbb{C}^N$.
Given this convention, we already know that $p \in \mathcal{B}_{\mu,t}$
by Theorem~\ref{thm1}, part~3.
Moreover, for any $p,q \in \mathcal{P}({\mathbb{R}^N})$ we have
\begin{equation}
\label{ipid}
   \langle \langle p, q \rangle \rangle_{\mu,t} = [p,q]_{\mu,t}.
\end{equation}
See \cite{SBSBO} and \cite{MR} for details in the case $t=1$.

We now take an orthonormal basis $\{\phi_\nu\}$ of $\mathcal{P}({\mathbb{R}^N})$
with respect to the inner product $[ \cdot , \cdot ]_{\mu,1}$.
To make things more specific we follow \cite{MR} by taking
$\nu \in \mathbb{Z}^N_+$  (that is, $\nu$ is a multi-index) and
$\phi_\nu$ to be a homogeneous polynomial of degree
$|\nu| = \nu_1 + \cdots + \nu_N$.
One can use the Gram-Schmidt procedure to show that such an
orthonormal basis exists.
Put $\phi_{t;\nu} = \delta_{t^{-1/2} } \phi_\nu$.
Then we have for $\nu, \pi \in \mathbb{Z}^N_+$ that
\begin{eqnarray*}
[ \phi_{t;\nu}, \phi_{t;\pi} ]_{\mu,t} &=& \langle e^{-t\Delta_\mu/2} \phi_{t;\nu} \, ,
e^{-t\Delta_\mu/2} \phi_{t;\pi} \rangle_{L^2(m_{\mu,t})}  \\
&=& \langle e^{-t\Delta_\mu/2} \delta_{t^{-1/2}} \phi_{\nu} \, ,
e^{-t\Delta_\mu/2} \delta_{t^{-1/2}} \phi_{\pi} \rangle_{L^2(m_{\mu,t})}  \\
&=& \langle \delta_{t^{-1/2}} e^{-\Delta_\mu/2} \phi_{\nu} \, ,
\delta_{t^{-1/2}} e^{-\Delta_\mu/2} \phi_{\pi} \rangle_{L^2(m_{\mu,t})}  \\
&=& \langle e^{-\Delta_\mu/2} \phi_{\nu} \, ,
e^{-\Delta_\mu/2} \phi_{\pi} \rangle_{L^2(m_{\mu,1})}  \\
&=& [ \phi_{\nu} \, ,\phi_{\pi} ]_{\mu,1}  \\
&=& \delta_{\nu, \pi}.
\end{eqnarray*}
Here we have used the identity
$e^{-t\Delta_\mu/2} \delta_{t^{-1/2}} = \delta_{t^{-1/2}} e^{-\Delta_\mu/2}$ which
holds, as the reader can check, basically because $\Delta_\mu$ is an operator
of degree $-2$.
So $\{ \phi_{t;\nu} \}$ is an orthonormal set in $\mathcal{P}({\mathbb{R}^N})$
with respect to the inner product $[ \cdot , \cdot ]_{\mu,t}$.
By using that any polynomial is in the span of $\{\phi_\nu\}$ and dilation,
it follows that $\{ \phi_{t;\nu} \}$ also spans $\mathcal{P}({\mathbb{R}^N})$
and so is also an orthonormal basis of $\mathcal{P}({\mathbb{R}^N})$.
Then Theorem~\ref{thm1}, part~3, and equation (\ref{ipid}) imply that $\{ \phi_{t;\nu} \}$ is
also an orthonormal basis of $\mathcal{B}_{\mu,t}$.

Next we define the {\em generalized Hermite polynomials} by
$$
H_{t;\nu} := e^{-t\Delta_\mu/2} \phi_{t;\nu}.
$$
It follows that $H_{t;\nu}$ is a polynomial of degree $|\nu|$, though not homogeneous in general.
We warn the reader that $H_{t;\nu}$ depends on $\mu$, though this is not shown in the notation.
Then we get
$$
  H_{t;\nu} = e^{-t\Delta_\mu/2} \phi_{t;\nu} =
  e^{-t\Delta_\mu/2} \delta_{t^{-1/2}} \phi_\nu = \delta_{t^{-1/2}}  e^{-\Delta_\mu/2} \phi_\nu =
  \delta_{t^{-1/2}}  H_{1;\nu},
$$
which shows that the generalized Hermite polynomials are dilations of the usual
generalized Hermite polynomials for $t=1$.
It also follows that
\begin{eqnarray*}
\langle H_{t;\kappa} , H_{t;\nu} \rangle_{L^2(m_{\mu,t} ) } &=&
\langle e^{-t\Delta_\mu/2} \phi_{t;\kappa}, e^{-t\Delta_\mu/2} \phi_{t;\nu} \rangle_{L^2(m_{\mu,t})} \\
&=& [ \phi_{t;\kappa} , \phi_{t,\nu} ]_{\mu,t} = \delta_{\kappa,\nu},
\end{eqnarray*}
which shows that $\{ H_{t;\nu} \}_\nu$ is an orthonormal set in $L^2(m_{\mu,t})$.
It can be shown that this is actually an orthonormal basis.
An important relation here is
\begin{equation}
\label{imprel}
  \sum_\nu H_{t;\nu} (x) \phi_{t;\nu} (y) =
        e^{-y^2/2t} E_\mu \left( \dfrac{x}{ t^{1/2} },\dfrac{y}{ t^{1/2} } \right)
\end{equation}
for $x \in \mathbb{C}^N$ and $y \in \mathbb{C}^N$.
Note that the polynomials $H_{t;\nu}$ depend on the choice of basis $\phi_\nu$.
One thing that we see in the previous identity is that the right side
does not depend on the choice of the basis $\{ \phi_\nu \}$ of $\mathcal{P}({\mathbb{R}^N})$.
Consequently, the left side of that identity also
does not depend on the choice of basis $\{ \phi_\nu \}$.

The {\em generalized Hermite functions} are then defined for $x \in \mathbb{R}^N$ by
$$
h_{t; \nu} (x) := e^{-x^2/4t} H_{t; \nu} (x).
$$
Note that the dependence of $h_{t; \nu} (x)$ on $\mu$ is not indicated in the notation.
Part of the importance of these functions lies in the next result.
\begin{prop}
The generalized Hermite functions $\{  h_{t; \nu}  \, | \, \nu \in \mathbb{Z}^N_+  \}$ are an
orthonormal basis of $L^2(\omega_{\mu,t})$.
\end{prop}
\textbf{Proof:}
To show this one simply notes that the transform
\begin{equation}
\label{groundstate}
  V_{\mu,t} \phi (x) :=\phi(x) ~/~ e^{-x^2/4t}
\end{equation}
defines a unitary isomorphism $V_{\mu,t} : L^2(\omega_{\mu,t}) \to L^2(m_{\mu,t}) $ and that
the inverse transform $V_{\mu,t}^{-1}$ maps the elements $H_{t;\nu}$
of an orthonormal basis of $ L^2(m_{\mu,t}) $ to the elements $h_{t;\nu}$ of $ L^2(\omega_{\mu,t})$.
$\quad \blacksquare$

We call $V_{\mu,t}$ the {\em ground state transformation}.
This is consistent with
the terminology in the case $\mu \equiv 0$, since $L^2(m_{\mu,t})$
uses a probability measure, has an orthonormal basis of Hermite type polynomials, and the
change of density function $e^{-x^2/4t}$ is associated with the eigenfunction
(unique up to constant multiple)
corresponding to the smallest eigenvalue of an appropriate generalization of the
harmonic oscillator Hamiltonian.

Next, we define the Version~A Segal-Bargmann transform in this context.
This can be defined as the integral kernel transform
\begin{equation}
\label{def-sbt}
 A_{\mu,t} \psi (z) :=
\int_{\mathbb{R}^N} \mathrm{d} \, \omega_{\mu,t}(q) A_{\mu,t} (z,q) \psi(q)
\end{equation}
for $\psi \in L^2 ( \mathbb{R}^N , \omega_{\mu,t})$ and $z \in \mathbb{C}^N$,
where the integral kernel is given by
\begin{equation}
\label{def-kernel-a}
A_{\mu,t} (z,q) := \exp \left(-z^2/2t -q^2/4t\right)
E_{\mu} \left( \dfrac{z}{ t^{1/2} },\dfrac{q}{ t^{1/2} }  \right)
\end{equation}
for $z \in \mathbb{C}^N$ and $q \in \mathbb{R}^N$.
We are here using the standard convention that the kernel function
and its associated integral kernel operator are represented by the same symbol,
which in this case is $A_{\mu,t}$.
We will continue using this convention in this paper.
Also note that $z^2$ is the holomorphic function $\sum_{j=1}^N z_j^2$
given that $z = ( z_1, \dots , z_N)$.

Before proceeding further, we should prove that the integral in (\ref{def-sbt})
makes sense.
So, writing $z=x+ i y$ with $x,y \in \mathbb{R}^N$, we estimate the kernel as follows:
\begin{eqnarray*}
| A_{\mu,t} (z,q) | &=& \exp \big( \!  -(x^2 -y^2)/2t -q^2/4t \big) \,
\Big| \, E_{\mu} \left( \dfrac{z}{ t^{1/2} },\dfrac{q}{ t^{1/2} }  \right) \Big| \\
&\le& \exp \big( \! -(x^2 -y^2)/2t -q^2/4t \big) \, \exp \left( |z| \, |q| / t \right),
\end{eqnarray*}
where we have used the estimate
\begin{equation}
\label{estEmu}
    |E_\mu(z,w)| \le \exp(|z| \, |w|)
\end{equation}
for $z,w \in \mathbb{C}^N$ and $\mu \ge 0$, which can be found in \cite{MR2}.
This clearly implies that $A_{\mu,t} (z,q)$ as a function of $q$ is in $L^2 ( \omega_{\mu,t})$.
But we are also taking $\psi \in L^2 ( \omega_{\mu,t})$.
So the integral in (\ref{def-sbt}) converges absolutely for all
$z \in \mathbb{C}^N$ by the Cauchy-Buniakowsky-Schwarz inequality.

Taking $N=1$ gives the Version~A Segal-Bargmann transform
as presented in \cite{SBS}.
For no value $t > 0$ is our definition identical to that
given in \cite{SBSBO}, though
it turns out our presentation is equivalent, as we shall show later on.
(See Corollary~\ref{big_cor}.)
While the formula for $A_{\mu,t} (z,x)$ was introduced here without
any motivation or further ado, the reader can find in \cite{SBSBO} an
enlightening way based on a \textit{restriction principle}
to construct those authors' equivalent transform.
However, let us note that the identity (\ref{imprel}) 
when multiplied by  $e^{-x^2/4t}$ implies immediately that
\begin{equation}
\label{Amut_id}
\sum_\nu h_{t;\nu} (x) \phi_{t;\nu} (z) = \exp \left(-z^2/2t -x^2/4t\right)
E_{\mu} \left( \dfrac{z}{ t^{1/2} },\dfrac{x}{ t^{1/2} }  \right) =
A_{\mu,t} (z, x)
\end{equation}
for $x \in \mathbb{R}^N$ and $z \in \mathbb{C}^N$.
The left side of this identity can be interpreted as the kernel function of an integral
transform that maps the orthonormal basis $\{  h_{t; \nu}(x) \}_\nu$ of $L^2(\omega_{\mu,t})$
to the the orthonormal basis $\{ \phi_{t;\nu}(z) \}_\nu$ of $\mathcal{B}_{\mu,t}$.
Therefore the Version~A Segal-Bargmann transform in this context maps
a ``canonical'' orthonormal basis to another ``canonical'' orthonormal basis by using
an integral kernel that is a standard infinite sum which just happens to be summable
in a closed form involving the Dunkl kernel function.

We now note that the Dunkl kernel is also used to define the Dunkl transform, a generalization
of the Fourier transform to this context.
This is defined for $\psi \in L^1 (\omega_{\mu,t})$ by
$$
   \mathcal{F}_{\mu,t} \psi (k) := \int_{\mathbb{R}^N} \mathrm{d}\omega_{\mu,t}(x)
   E_\mu \left( \dfrac{-ik}{t^{1/2}}, \dfrac{x}{t^{1/2}} \right) \psi (x),
$$
where $t>0$ and $k \in \mathbb{R}^N$.
This transform can be extended to $L^2 (\omega_{\mu,t})$, and it then becomes
a unitary isomorphism of the Hilbert space $L^2 (\omega_{\mu,t})$ to itself,
and so we say that the measure $\mathrm{d} \omega_{\mu,t} $ is \textit{self dual}.
This transform diagonalizes simultaneously all the Dunkl operators $T_{\xi,\mu} $,
where $\xi \in \mathbb{R}^N$,
according to the formula
\begin{equation}
\label{FstarTF}
        \mathcal{F}_{\mu,t}^* T_{\xi,\mu} \mathcal{F}_{\mu,t} = -\dfrac{i}{t} M_{\xi},
\end{equation}
where $M_{\xi} \psi (x) := \langle \xi, x \rangle \psi(x)$
for $x \in \mathbb{R}^N$ and the adjoint of the Dunkl operator is given by
$$
   \mathcal{F}_{\mu,t}^* \phi (x) := \int_{\mathbb{R}^N} \mathrm{d}\omega_{\mu,t}(k)
   E_\mu \left( \dfrac{ik}{t^{1/2}}, \dfrac{x}{t^{1/2}} \right) \phi (k).
$$
Notice that this diagonalizes the family of Dunkl operators $T_{\xi,\mu}$ for $\xi \in \mathbb{R}^N$
(which does not depend  on $t$) for any $t>0$.
However, the resulting multiplication operators on the right side of (\ref{FstarTF}) do depend on $t$.
Of course, this simultaneous diagonalization of the family $T_{\xi,\mu}$ implies that
this is a family of commuting unbounded operators acting in the Hilbert space $L^2 (\omega_{\mu,t})$.
(Note that $M_\xi$ is not bounded if $\xi \ne 0$.)

Similarly, we have
\begin{equation}
\label{FstarMF}
       \mathcal{F}_{\mu,t}^* M_{\xi} \mathcal{F}_{\mu,t} = -i t  T_{\xi,\mu}.
\end{equation}

The formulas (\ref{FstarTF}) and (\ref{FstarMF}) can also be written in terms
of the \textit{Dunkl momentum operators} $P_{\xi,\mu,t} := (t/i) T_{\xi,\mu}$ as
\begin{gather*}
  \mathcal{F}_{\mu,t}^* P_{\xi,\mu,t} \mathcal{F}_{\mu,t} = - M_{\xi}, \\
   \mathcal{F}_{\mu,t}^* M_{\xi} \mathcal{F}_{\mu,t} =  P_{\xi,\mu,t},
\end{gather*}
showing that $M_{\xi,\mu,t}:=M_{\xi}$ should be called a \textit{Dunkl position operator}.
The commutation relations between the Dunkl momentum and Dunkl position
operators are known, but will not be needed here.
(See \cite{SBSBO} and \cite{SO}.)

Next we generalize the definition of translation operator to this context.
See \cite{RO} for the case $N = 1$ and \cite{MR} for the case of general finite $N$.
A good reference for this translation operator as well as for its convolution operator
(which we will discuss next) is \cite{TX}.

\begin{definition}
The {\em Dunkl translation operator by  $y \in \mathbb{R}^N$}, denoted $\mathcal{T}_{\mu,y}$,
is defined by
$$
\mathcal{T}_{\mu,y} : = E_\mu \left( -y , T_\mu \right),
$$
where $T_\mu$ is the {\em Dunkl gradient} defined by
$$
          T_\mu := ( T_{\epsilon_1,\mu}, \dots,  T_{\epsilon_j,\mu}, \dots ,   T_{\epsilon_N,\mu})
$$
and $\epsilon_1, \dots , \epsilon_j, \dots , \epsilon_N$ is the standard basis
of $\mathbb{R}^N$.
\end{definition}
To be more specific we need to define functions of the commuting family of operators
$\{ T_{\epsilon_j,\mu}  \}_{j=1}^N $.
This can be done with spectral theory, of course.
But we have already seen that the Dunkl transform gives a simultaneous diagonalization
of this family.
So, using (\ref{FstarMF}), we have for any $t>0$ that
\begin{equation}
\label{def_mu_trans}
\mathcal{T}_{\mu,y} =
\mathcal{F}_{\mu,t}^* E_\mu \left( -y, \dfrac{ik}{t} \right) \mathcal{F}_{\mu,t},
\end{equation}
where $k \in \mathbb{R}^N$ denotes the independent variable
in the domain space of $\mathcal{F}_{\mu,t}^*$.
Notice that the left side of (\ref{def_mu_trans}) is independent of $t$.

Since $| E_\mu ( -y, ik/t )| \le 1$
for $ \mu \ge 0$ (see \cite{MR}), we have that $\mathcal{T}_{\mu,y}$ is a bounded operator
on $L^2(\omega_{\mu,t}) $ whose norm is bounded by $1$.
For $\phi \in L^2(\omega_{\mu,t})$ this defines
$\mathcal{T}_{\mu,y} \phi$ as an element in $L^2(\omega_{\mu,t})$
and so $\mathcal{T}_{\mu,y} \phi (x)$ for
$\omega_{\mu,t}$-almost all $x \in \mathbb{R}^N$.

We can expand the formula (\ref{def_mu_trans}) into an iterated integral as follows:
\begin{gather*}
\mathcal{T}_{\mu,y} \phi (x) = \left(
\mathcal{F}_{\mu,t}^* E_\mu \left( -y, \dfrac{ik}{t} \right) \mathcal{F}_{\mu,t} \right) \phi (x)
\\
= \int_{\mathbb{R}^N}
 \mathrm{d} \omega_{\mu,t}(k) E_\mu \left( \dfrac{k}{t^{1/2}},\dfrac{ix}{t^{1/2}}
 \right)
E_\mu \left(  -\dfrac{y}{t^{1/2}}, \dfrac{ik}{t^{1/2}} \right) \mathcal{F}_{\mu,t}\phi(k)
\\
=  \int_{\mathbb{R}^N}
 \mathrm{d} \omega_{\mu,t}(k) E_\mu \left( \dfrac{k}{t^{1/2}},\dfrac{ix}{t^{1/2}}
 \right)
 E_\mu \left( -\dfrac{y}{t^{1/2}}, \dfrac{ik}{t^{1/2}} \right) \cdot \\
\int_{\mathbb{R}^N}
 \mathrm{d} \omega_{\mu,t} (q) E_\mu \left( -\dfrac{ik}{t^{1/2}},\dfrac{q}{t^{1/2}} \right) \phi(q).
\end{gather*}
Note that our definition differs by a sign from those given in \cite{RO} and \cite{MR}.
In particular, if $\mu \equiv 0$, then $\mathcal{T}_{0,y} \phi (x) = \phi(x-y)$
with our definition.

Also one can use the power series expansion of the holomorphic function
$E_\mu : \mathbb{C}^N \times \mathbb{C}^N \to \mathbb{C}$ 
to define the function $E_\mu \left( -y , T_\mu \right) \phi (x) $ pointwise
in $x \in \mathbb{R}^N$, but we do not wish to go into details.

Now that we have defined a translation operator, a natural next step is
to define an associated convolution operator as has been done in \cite{BE}
and \cite{SBS} in dimension $N=1$.
The case of arbitrary finite dimension is treated in \cite{TX}.

\begin{definition}
For functions $\phi,\psi: \mathbb{R}^N \to \mathbb{C}$ and $x \in \mathbb{R}^N$
we define their {\em Dunkl convolution product} $\ast_{\mu,t}$ by
$$
\left(\phi \ast_{\mu,t} \psi \right) (x): = \int_{\mathbb{R}^N} \mathrm{d} \omega_{\mu,t}(y) \, (\mathcal{T}_{\mu,y} \phi) (x) \, \psi(y),
$$
provided that
$\mathcal{T}_{\mu,y} \phi (x)$ is defined for
$\omega_{\mu,t}$-almost all $y \in \mathbb{R}^N$ and
that the integral converges absolutely.
\end{definition}

When $\mu\equiv0$ and $t=1$, this definition reduces to the
standard convolution operation of classical analysis.
Also, if $N=1$, $\mu \in (-1/2,\infty)$
and $t=1$, this coincides with the definition of $\mu$-deformed convolution
given in \cite{SBS}.

Finally, we present the heat kernel associated with this theory.
(See R\"osler's papers \cite{MR} and \cite{MR2}.)
The heat equation in this theory is
$$
 \dfrac{\partial u}{\partial t} = \dfrac{1}{2} \Delta_\mu u,
$$
where the solution is a suitably smooth function
$u: \mathbb{R}^N \times [0,\infty) \to \mathbb{R}$.
Also, $\Delta_\mu$ is the Dunkl Laplacian introduced earlier.
It is at this point that great attention must be put to the problem
of identifying correctly the formula for the heat kernel, since none of
the conventions in use (as far as we are aware) agree with our conventions.
So, given our conventions, the kernel function for the associated initial value problem is
$$
\rho_{\mu,t} (x,q) =
e^{-(x^2 + q^2)/2t} E_\mu \left( \dfrac{x}{ t^{1/2} }, \dfrac{q}{ t^{1/2} } \right)
$$
and is known as the {\em (Dunkl) heat kernel}.
Here $x,q \in \mathbb{R}^N$.
The point is that the initial value problem is solved by
$$
u(x,t) = \int_{\mathbb{R}^N} \mathrm{d} \omega_{\mu,t} (q) \rho_{\mu,t} (x, q) f(q),
$$
where $u(x,0) = f(x)$ is the initial condition.
(Of course, we are omitting some technical hypotheses.)
We are associating the factor $c_\mu^{-1} t^{-(\gamma_\mu + N/2)}$ that appears
in the usual formula for the heat kernel with the measure
$\mathrm{d} \tilde{\omega}_{\mu,1}$ to give us the measure $\mathrm{d} \omega_{\mu,t}$.
(Again see \cite{MR} and \cite{MR2}.)
It will undoubtedly look strange to some readers to have a measure, as well as a heat kernel,
depending on the time parameter $t$.
However, in our defense, it is a way to make the rest of the theory ``work out.''
Besides, it is mathematically correct and rigorous.
But we wish to emphasize that in our convention the Dunkl heat kernel $\rho_{\mu, t}$ is
\textit{not} a solution of the Dunkl heat equation.

Moreover, note that the heat equation here has a factor of $1/2$ that is
not present in \cite{MR} and \cite{MR2}, and this
leads to a change in the time parameter in our formula for the heat kernel.
Also, note that $\rho_{\mu,t} : \mathbb{R}^N \times \mathbb{R}^N \to \mathbb{R}$
admits a holomorphic continuation $\mathbb{C}^N \times \mathbb{C}^N \to \mathbb{C}$,
which we also denote by $\rho_{\mu,t}$.

\section{Results}
\label{sec3}

We would like to show explicitly that the Segal-Bargmann
spaces $\mathcal{B}_{\mu,t}$, where $t>0$, are all unitarily equivalent in a natural way.
We will do this in terms of dilation operators.
We already have remarked that
$\delta_\lambda : L^2( \omega_{\mu,t}) \to L^2( \omega_{\mu,s})$
is a unitary isomorphism, provided that $ \lambda = (t/s)^{1/2}$.
The corresponding result for the Segal-Bargmann
spaces can not use a change of variables argument, since the
inner product is not defined in terms of a measure.
So we give its proof next.

\begin{prop}
\label{p1}
Suppose that $t>0$ and $s>0$ are given.
Then for any multiplicity function $\mu \ge 0$ we have that
the dilation operator $D_{(t/s)^{1/2}}$ is a unitary isomorphism
$\mathcal{B}_{\mu,t} \to \mathcal{B}_{\mu,s}$.
\end{prop}
{\bf Proof:}
A unitary operator preserves the inner product.
However, the inner product
$\langle\langle \cdot \, , \cdot \rangle\rangle_{\mu,t}$ is not
given as an $L^2$ type inner product with respect to some measure on
$\mathbb{C}^N$, but rather in terms of inner products of functions
of the form $\mathbb{K}_{z,\mu,t}$. (See Theorem 3.2 in \cite{SBSBO}.)
So, we begin with a calculation of the inner product of two
functions of that form. We have that
\begin{eqnarray*}
\langle\langle \mathbb{K}_{z,\mu,t} \, , \mathbb{K}_{w,\mu,t}
\rangle\rangle_{\mu,t} &=& \mathbb{K}_{w,\mu,t}(z) \\
&=&\mathbb{K}_{\mu,t}(w,z) \\
&=&E_\mu \left( \dfrac{w^*}{ t^{1/2} }, \dfrac{z}{ t^{1/2} } \right),
\end{eqnarray*}
where the first equality is the reproducing kernel property,
the second is the definition of $\mathbb{K}_{w,\mu,t}$ and the third is the
definition of $\mathbb{K}_{\mu,t}$.
Replacing $t$ with $s>0$ in this, we obtain
$$
\langle\langle \mathbb{K}_{z,\mu,s} \, , \mathbb{K}_{w,\mu,s}
\rangle\rangle_{\mu,s} = E_\mu \left( \dfrac{w^*}{ s^{1/2} }, \dfrac{z}{ s^{1/2} } \right),
$$
which implies immediately that
$$
\langle\langle \mathbb{K}_{(s/t)^{1/2}z,\mu,s} \, ,
 \mathbb{K}_{(s/t)^{1/2}w,\mu,s}
\rangle\rangle_{\mu,s} = E_\mu \left( \dfrac{w^*}{ t^{1/2} }, \dfrac{z}{ t^{1/2} } \right).
$$

Since the finite linear combinations of the functions
$\mathbb{K}_{z,\mu,t}$ are dense
in $\mathcal{B}_{\mu,t}$ (see \cite{SBSBO}), we have
a unitary isomorphism of Hilbert spaces
$U_{t,s} : \mathcal{B}_{\mu,t} \to \mathcal{B}_{\mu,s}$
given by
$$
U_{t,s} : \mathbb{K}_{z,\mu,t} \mapsto \mathbb{K}_{(s/t)^{1/2}z,\mu,s}.
$$
The inverse map is clearly $U_{s,t}$.
Now we write this in terms of a dilation operator.
One just uses the (yet unproved) identity
\begin{equation}
\label{dkk}
D_{(t/s)^{1/2}} \mathbb{K}_{z,\mu,t}
 = \mathbb{K}_{(s/t)^{1/2}z,\mu,s}.
\end{equation}
This will give us immediately the formula
$U_{t,s} = D_{(t/s)^{1/2}}$ and thereby proves the theorem.
To prove the identity (\ref{dkk}), we start with the left side.
For all $w \in \mathbb{C}^N$ we have
\begin{eqnarray*}
 D_{(t/s)^{1/2}} \mathbb{K}_{z,\mu,t} (w) &=&
 \mathbb{K}_{z,\mu,t}((t/s)^{1/2}w) \\
&=& \mathbb{K}_{\mu,t}(z, (t/s)^{1/2}w) \\
&=& E_\mu \left( \dfrac{z^*}{ t^{1/2} } , \dfrac{1}{ t^{1/2} } (t/s)^{1/2}w \right) \\
&=& E_\mu \left( \dfrac{z^*}{ t^{1/2} } , \dfrac{w}{ s^{1/2} } \right).
\end{eqnarray*}
Next, the right side of (\ref{dkk}) gives us
\begin{eqnarray*}
\mathbb{K}_{(s/t)^{1/2}z,\mu,s}(w) &=& \mathbb{K}_{\mu,s}((s/t)^{1/2}z,w)
\\
&=& E_\mu \left( \dfrac{1}{ s^{1/2} } (s/t)^{1/2}z^* , \dfrac{w}{ s^{1/2} } \right)
\\
&=& E_\mu \left( \dfrac{z^*}{ t^{1/2} } , \dfrac{w}{ s^{1/2} } \right)
\end{eqnarray*}
for all $w \in \mathbb{C}^N$.
This proves  (\ref{dkk}).
$\quad \blacksquare$

\vskip .5cm

We now have a result showing the relation between the Segal-Bargmann
transforms $A_{\mu,t}$ and $A_{\mu,s}$.

\begin{theorem}
\label{comm_diagram}
Let $s$ and $t$ be positive numbers.
Suppose that the multiplicity function is non-negative, that is,
$\mu : \mathcal{R} \to [0,\infty)$.
Then the diagram
\begin{equation}
\begin{array}{ccc}
L^2(\omega_{\mu,t}) & \stackrel{\delta_{(t/s)^{1/2} } }   {\longrightarrow} & L^2(\omega_{\mu,s}) \\
\Big\downarrow\vcenter{ \llap{ $A_{\mu,t}~~$ }  }  & & 
\Big\downarrow\vcenter{ \rlap{ $A_{\mu,s}$ }  } \\
\mathcal{B}_{\mu,t} & \stackrel{\longrightarrow}{D_{(t/s)^{1/2} } } & \mathcal{B}_{\mu,s}
\end{array}
\end{equation}
commutes.
\end{theorem}
{\bf Proof:} We take a function $\psi \in L^2(\omega_{\mu,t})$
in the upper left corner and chase it through the diagram.
First, we go down and across.
Using definitions we get for all $z \in \mathbb{C}^N$ that
\begin{eqnarray*}
  && ( D_{(t/s)^{1/2}} A_{\mu,t} \psi ) (z) =
         A_{\mu,t} \psi \left( (t/s)^{1/2} z \right) \\
&=& \int_{\mathbb{R}^N } \mathrm{d}^N x \, \omega_{\mu,t} (x)
A_{\mu,t}  \left( (t/s)^{1/2} z , x \right) \psi(x) \\
&=& \int_{\mathbb{R}^N } \mathrm{d}^N x \, \omega_{\mu,t} (x)
\exp \left(-\dfrac{z^2}{2t} \dfrac{t}{s} - \dfrac{x^2}{4t} \right)
E_{\mu} \left( \left( \dfrac{t}{s} \right)^{1/2}
 \dfrac{z}{ t^{1/2} },\dfrac{x}{ t^{1/2} }  \right) \psi(x) \\
&=& \int_{\mathbb{R}^N } \mathrm{d}^N x \, \omega_{\mu,t} (x)
e^{-z^2/2s} e^{-x^2/4t}
E_{\mu} \left( \dfrac{z}{ s^{1/2}},\dfrac{x}{ t^{1/2} }  \right) \psi(x).
\end{eqnarray*}

Now we go across and then down.
Using definitions and the change of variable $y =(t/s)^{1/2}x$
for the third equality, we obtain for all $z \in \mathbb{C}^N$ that

\begin{eqnarray*}
&&\left( A_{\mu,s} \delta_{(t/s)^{1/2} } \psi\right) (z)\\
&=& \int_{\mathbb{R}^N } \mathrm{d}^N x \, \omega_{\mu,s} (x)
A_{\mu,s}  \left( z , x \right) ( \delta_{(t/s)^{1/2} } \psi) (x) \\
&=& \int_{\mathbb{R}^N } \mathrm{d}^N x \, \omega_{\mu,s} (x)
A_{\mu,s}  \left( z , x \right)
\psi ( ( t/s )^{1/2} x ) \\
&=& \int_{\mathbb{R}^N } \mathrm{d}^N y \,
\left( \dfrac{t}{s} \right)^{-(\gamma_\mu+N/2)} \omega_{\mu,s} (y)
A_{\mu,s}  \left( z , \left(\dfrac{s}{t}\right)^{1/2} y \right)
\psi(y)\\
&=&\int_{\mathbb{R}^N } \mathrm{d}^N y \, \, \omega_{\mu,t} (y)
A_{\mu,s}  \left( z , \left(\dfrac{s}{t}\right)^{1/2} y \right) \psi(y)\\
&=& \int_{\mathbb{R}^N } \mathrm{d}^N y \, \omega_{\mu,t} (y)
\exp \left( -\dfrac{z^2}{2s} - \, \dfrac{s}{t} \dfrac{y^2}{4s} \right)
E_{\mu}\left( \dfrac{z}{ s^{1/2} }, \left(\dfrac{s}{t}\right)^{1/2}
\dfrac{y}{ s^{1/2}  } \right) \psi(y) \\
&=& \int_{\mathbb{R}^N } \mathrm{d}^N y \, \omega_{\mu,t} (y)
e^{-z^2/2s} e^{-y^2/4t}
E_{\mu}\left( \dfrac{z}{ s^{1/2} }, \dfrac{y}{ t^{1/2} } \right) \psi(y).
\end{eqnarray*}
So the diagram commutes as claimed. $\quad \blacksquare$

The following theorem has been proved in \cite{SBSBO} and \cite{SO}
in the case $t=1$.
Since our normalizations are different from theirs and since we explicitly
use Planck's constant, we prove it here.
\begin{theorem}
Suppose that $t>0$ and that the multiplicity function satisfies $\mu \ge 0$.
Then the Segal-Bargmann transform
$A_{\mu,t}: L^2(\omega_{\mu,t}) \to \mathcal{B}_{\mu,t}$ is a unitary
isomorphism.
\end{theorem}
{\bf Proof:}
Since this is a central result in this theory, we offer two proofs.
We will first show that the Segal-Bargmann transform of \cite{SBSBO}
can be factorized as the product of three maps.
Specifically we consider the composite map
$$
L^2( \mathbb{R}^N, \, \tilde{\omega}_{\mu,1}) \stackrel{R}{\longrightarrow}
 L^2( \mathbb{R}^N, \, \omega_{\mu,1/4}) \stackrel{\delta_{1/2} } {\longrightarrow}
L^2( \mathbb{R}^N, \, \omega_{\mu,1})
\stackrel{A_{\mu,1}}{\longrightarrow} \mathcal{B}_{\mu,1},
$$
where the first arrow is the map $R: \phi \mapsto c_\mu^{1/2} \, 2^{-(\gamma_\mu + N/2)} \phi$.
We claim that the rescaling map $R$ is a unitary isomorphism.
To show this we first note that
\begin{eqnarray*}
&& \langle R \phi_1, R \phi_2 \rangle_{L^2(\omega_{\mu,1/4})} =
 \langle c_\mu^{1/2} \, 2^{-(\gamma_\mu + N/2)} \phi_1,
c_\mu^{1/2} \, 2^{-(\gamma_\mu + N/2)} \phi_2 \rangle_{L^2(\omega_{\mu,1/4})}
\\
&=& \int_{\mathbb{R}^N} \mathrm{d} \omega_{\mu,1/4 } \, (x) \,
 \, c_\mu \, 2^{-2(\gamma_\mu + N/2)} \phi_1^*(x) \phi_2 (x)
\\
&=& \int_{\mathbb{R}^N} \mathrm{d} \tilde{\omega}_{\mu,1}(x) c_\mu^{-1} \,
 \left(\dfrac{1}{4}\right)^{-(\gamma_\mu + N/2)}
 c_\mu \, 2^{-2(\gamma_\mu + N/2)} \phi_1^*(x) \phi_2 (x)
\\
&=& \int_{\mathbb{R}^N} \mathrm{d} \tilde{\omega}_{\mu,1}(x) \,
\phi_1^*(x) \phi_2 (x) \\
&=& \langle \phi_1, \phi_2 \rangle_{L^2(\tilde{\omega}_{\mu,1})}.
\end{eqnarray*}
This shows that $R$ is unitary.
But $R$ is clearly invertible, and so is a unitary isomorphism.

We now compute the composite of the three maps given above.
Using definitions, some identities and the change of
variable $y = x/2$ in the fourth equality, we have for
$ \phi \in L^2(\tilde{\omega}_{\mu,1})$ and $z \in \mathbb{C}^N$ that
\begin{eqnarray*}
&& (A_{\mu,1} \, \delta_{1/2} \, R \phi) (z) = \int_{\mathbb{R}^N}\mathrm{d} \omega_{\mu,1}(x)
A_{\mu,1}(z,x) \left(  \delta_{1/2} \, R \phi \right) (x)
\\
&=& \int_{\mathbb{R}^N}\mathrm{d} \omega_{\mu,1}(x)
A_{\mu,1}(z,x) \left( R \phi \right) (x/2)\\
&=& \int_{\mathbb{R}^N}\mathrm{d} \omega_{\mu,1}(x)
A_{\mu,1}(z,x) \, (c_\mu^{1/2} \,  2^{-(\gamma_\mu + N/2)}) \phi (x/2)
\end{eqnarray*}
\begin{eqnarray*}
&=& \int_{\mathbb{R}^N}\mathrm{d} \tilde{\omega}_{\mu,1}(y) \, c_\mu^{-1} \, 2^{2\gamma_\mu + N}
A_{\mu,1}(z,2y) (c_\mu^{1/2} 2^{-(\gamma_\mu + N/2)}) \phi (y)
\\
&=& \int_{\mathbb{R}^N}\mathrm{d} \tilde{\omega}_{\mu,1}(y) 2^{\gamma_\mu + N/2}
c_\mu^{-1/2} \exp \left( -z^2/2 - y^2 \right) E_\mu (z, 2y) \phi (y).
\end{eqnarray*}
Using the identity $E_\mu (z, 2y) = E_\mu (\sqrt{2} y, \sqrt{2} z)$, we see that this composite
is an integral kernel operator $ BSO :L^2(\tilde{\omega}_{\mu,1}) \to  \mathcal{B}_{\mu,1}$
whose kernel for $z \in \mathbb{C}^N$ and $y \in \mathbb{R}^N$ is
$$
BSO(z,y): =
2^{\gamma_\mu + N/2} c_\mu^{-1/2} \exp \left( -z^2/2 - y^2 \right) E_\mu (\sqrt{2} y, \sqrt{2} z).
$$
(We conventionally do not include any part of the measure in the kernel function of an
integral kernel operator.)
Note that this is now exactly the kernel function for the Segal-Bargmann transform in \cite{SBSBO},
Theorem 4.2.
But Ben~Sa\"id and {\O}rsted show in \cite{SBSBO} that their
Segal-Bargmann transform $BSO$ is a unitary isomorphism
from  $L^2(\tilde{\omega}_{\mu,1})$ onto $\mathcal{B}_{\mu,1}$ (in our notation).
So the composite $A_{\mu,1} \, \delta_{1/2} \, R$ is a unitary isomorphism.
But we already know that $\delta_{1/2}$ and $R$ are unitary isomorphisms.
It follows that $A_{\mu,1}$ is also a unitary isomorphism.
But then we use the commutative diagram in
Theorem \ref{comm_diagram} (with $s=1$) to show that $A_{\mu,t}$
is a unitary isomorphism for any $t>0$.

For the second proof we simply note again that identity (\ref{Amut_id}) shows that
$A_{\mu,t}$ maps the orthonormal basis $\{ h_{t;\nu} \}$
of $L^2(\omega_{\mu,t})$ to the orthonormal basis $\{ \phi_{t;\nu} \}$
of $\mathcal{B}_{\mu,t}$.
And that implies that $A_{\mu,t}$ is a unitary isomorphism. $\quad \blacksquare$

An important aspect of the first proof of this theorem is that it
shows how the Segal-Bargmann transform defined in this article is related to the
Segal-Bargmann transform of \cite{SBSBO}.
We state next this corollary of the proof.

\begin{corollary}
\label{big_cor}
The Segal-Bargmann transform $BSO$ defined by
Ben~Sa\"id and {\O}rsted in \cite{SBSBO} and the Segal-Bargmann transform
$A_{\mu,1}$ of this article are related by
the formula
$$
BSO = A_{\mu,1} \, \delta_{1/2} \, R,
$$
where $R$ is the rescaling map defined above.
So $BSO$ and $A_{\mu,1}$ differ by a unitary isomorphism which is
a combination of a rescaling and a dilation on the configuration space $\, \mathbb{R}^N$.
This constitutes the rigorous assertion behind the statement that  $BSO$ and $A_{\mu,t}$
for $t=1$ are essentially the same transform.
\end{corollary}

We now enter into a topic that is based on results of Hall in \cite{HA},
where the original Euclidean case of the Segal-Bargmann
transform is treated.
This was generalized to the case of $\mu$-deformed quantum mechanics
in \cite{SBS}.
These two cases are now themselves generalized in the following theorems to the
present context.
\begin{theorem}
\label{version_a}
(Version A)
Suppose  that the multiplicity function satisfies $\mu \ge 0$ and that $\, t>0$.
The kernel of the Version~A Segal-Bargmann transform is then related to the
Dunkl heat kernel by the identity
\begin{equation}
\label{A_rho_id}
  A_{\mu,t} (z,q) = \frac{\rho_{\mu,t}(z,q)}{(\rho_{\mu,t}(0,q) )^{1/2}}
\end{equation}
for all $z \in \mathbb{C}^N$ and $q \in \mathbb{R}^N$.
\end{theorem}
{\bf Proof:}
The function $\rho_{\mu,t}$ in the numerator on the right side of equation (\ref{A_rho_id}) is the
analytic continuation of the Dunkl heat kernel, which does exist as we already have noted.
Next, we note that
$$
\rho_{\mu,t}(0,q) = e^{-q^2/2t}.
$$
Since this is a strictly positive function, the square root in the
denominator on the right hand side of the identity (\ref{A_rho_id})
is taken to be the positive square root.
Then we calculate
\begin{eqnarray*}
\frac{\rho_{\mu,t}(z,q)}{(\rho_{\mu,t}(0,q) )^{1/2}} &=& e^{-(z^2 + q^2)/2t}
E_\mu \left( \dfrac{z}{ t^{1/2} }, \dfrac{q}{ t^{1/2} } \right) e^{q^2/4t} \\
&=& e^{-z^2/2t} e^{-q^2/4t}
E_\mu \left( \dfrac{z}{ t^{1/2} }, \dfrac{q}{ t^{1/2} } \right)  \\
&=& A_{\mu,t} (z,q).
\end{eqnarray*}
And this shows the identity (\ref{A_rho_id}). $\quad \blacksquare$

Note that if $N=1$, then this identity reduces to an identity of the same form
in \cite{SBS} for the case of $\mu$-deformed
quantum mechanics considered there.
Also, in the case of arbitrary finite $N \ge 1$ and $\mu \equiv 0$
this identity reduces to the Euclidean case of dimension $N$ as given
in \cite{HA}.
In fact, the relation (\ref{A_rho_id}) is identical in form to the relations
given in \cite{HA} and \cite{SBS}, though the normalizations here are different.

There is a reformulation of the previous theorem in terms of the ground state Hilbert space,
$ L^2(\mathbb{R}^N, m_{\mu,t})$.
This is Version~B of the theory.

\begin{theorem}
\label{version_b}
(Version B) Define a kernel function by
$$
 B_{\mu,t} (z,q) := \frac{\rho_{\mu,t}(z,q)}{\rho_{\mu,t}(0,q) }
$$
for $z \in \mathbb{C}^N$ and $q \in \mathbb{R}^N$
and its associated integral kernel transform, which
is called the {\em Version~B Segal-Bargmann transform}, by
\begin{equation}
\label{def_B_mu_t}
 B_{\mu,t} \phi (z) :=
  \int_{ \mathbb{R}^N } \mathrm{d} m_{\mu,t}(q) B_{\mu,t} (z,q) \phi(q)
\end{equation}
for $ \phi \in L^2(\mathbb{R}^N, m_{\mu,t})$ and $z \in \mathbb{C}^N$.
Then
 $B_{\mu,t} : L^2(\mathbb{R}^N, m_{\mu,t}) \to \mathcal{B}_{\mu,t}$
is a unitary isomorphism.
\end{theorem}
{\bf Proof:}
A simple estimate using (\ref{estEmu})
shows that $ B_{\mu,t} (z,\cdot) \in L^2(\mathbb{R}^N, m_{\mu,t}) $
and so the integral in (\ref{def_B_mu_t}) converges absolutely for every $z \in \mathbb{C}^N$.

Since each map in the diagram
$$
 L^2(m_{\mu,t}) \stackrel{V_{\mu,t}^{-1}}{\longrightarrow} L^2(\omega_{\mu,t})
 \stackrel{A_{\mu,t}}{\longrightarrow} \mathcal{B}_{\mu,t}
$$
is a unitary isomorphism, then so is their composition.
We recall that the unitary transform $V_{\mu, t} $ was defined in equation (\ref{groundstate}).
We claim that this composition is an integral kernel transform whose
kernel function is precisely $B_{\mu,t} (z,q)$ as defined above.
So for $ \phi \in L^2(\mathbb{R}^N, m_{\mu,t})$ we calculate
\begin{eqnarray*}
( A_{\mu,t} V_{\mu,t}^{-1} \phi) (z) &=& \int_{\mathbb{R}^N} \mathrm{d} \omega_{\mu,t} (q)
A_{\mu,t} (z,q) V_{\mu,t}^{-1} \phi (q) \\
&=& \int_{\mathbb{R}^N} \mathrm{d} \omega_{\mu,t} (q)
A_{\mu,t} (z,q) e^{-q^2/4t} \phi(q) \\
&=& \int_{\mathbb{R}^N} \mathrm{d} \omega_{\mu,t} (q)
\left( \dfrac{\rho_{\mu,t}(z,q)}{(\rho_{\mu,t}(0,q) )^{1/2}} \right)
 \left( \rho_{\mu,t}(0,q) \right)^{1/2} \phi(q) \\
&=& \int_{\mathbb{R}^N} \mathrm{d} \omega_{\mu,t} (q) \rho_{\mu,t}(z,q) \phi(q) \\
&=& \int_{\mathbb{R}^N} \mathrm{d} \omega_{\mu,t} (q) \rho_{\mu,t}(0,q)
\left( \dfrac{\rho_{\mu,t}(z,q)}{\rho_{\mu,t}(0,q)} \right) \phi(q) \\
&=& \int_{\mathbb{R}^N} \mathrm{d} m_{\mu,t} (q)
\left( \dfrac{\rho_{\mu,t}(z,q)}{\rho_{\mu,t}(0,q)} \right) \phi(q) \\
&=& \int_{\mathbb{R}^N} \mathrm{d} m_{\mu,t} (q) B_{\mu,t} (z,q) \phi(q),
\end{eqnarray*}
where in the penultimate step we used the identity
$$
 \mathrm{d} m_{\mu,t} (q) = e^{-q^2/2t} \mathrm{d} \omega_{\mu,t} (q)
  = \rho_{\mu,t} (0,q)  \mathrm{d} \omega_{\mu,t} (q).
$$
Note that the expression after the fourth equality sign does {\em not} tell
us that the kernel is $\rho_{\mu,t}(z,q)$.
But the last two expressions, which are integrals with respect to the measure
$\mathrm{d} m_{\mu,t}$ of the domain $L^2$ space, do give us the kernel
function.

So the integral kernel transform $B_{\mu, t}$ is a unitary isomorphism, since it is equal
to the unitary isomorphism $A_{\mu,t} V_{\mu,t}^{-1}$. $\quad \blacksquare$

As with the previous theorem, in the case $N=1$ this
result reduces to a result
in \cite{SBS}, while for $N \ge 1$ and $\mu \equiv 0$
we get the original result of this type given in \cite{HA}.
And again we get here exactly the same formulas as found
in \cite{HA} and \cite{SBS} modulo normalizations.

Next we will discuss Version~C of this Segal-Bargmann analysis.
The notion of a Version~C was originally introduced in \cite{HA} in the context of
the Segal-Bargmann analysis of compact Lie groups and related differential manifolds.
First, we give some definitions needed for this.

\begin{definition}
Define the kernel function
$$
 C_{\mu,t} (z,q) := \rho_{\mu,t}(z,q)
$$
for $z \in \mathbb{C}^N$ and $q \in \mathbb{R}^N$
and its associated integral kernel transform, which
is called the {\em Version~C Segal-Bargmann transform},
\begin{equation}
\label{def_C_mu_t}
 C_{\mu,t} \phi (z) :=
  \int_{ \mathbb{R}^N } \mathrm{d} \omega_{\mu,t}(q) C_{\mu,t} (z,q) \phi(q)
\end{equation}
for $ \phi \in L^2(\mathbb{R}^N, \omega_{\mu,t})$.
(A straightforward estimate using (\ref{estEmu}) shows that
$ C_{\mu,t} (z,\cdot) \in L^2(\mathbb{R}^N, \omega_{\mu,t}) $
and so the integral in (\ref{def_C_mu_t}) converges absolutely for every $z \in \mathbb{C}^N$.)

Also, we write $\mathcal{H} (\mathbb{C}^N)$ for the complex vector space of all
holomorphic functions $f: \mathbb{C}^N \to \mathbb{C}$.
For $f \in \mathcal{H} (\mathbb{C}^N)$ we define $Gf \in \mathcal{H} (\mathbb{C}^N)$
by
$$
         Gf(z) := 2^{\gamma_\mu/2 + N/4} f(2z)/A_{\mu,2t}(2z,0)
$$
for $z \in \mathbb{C}^N$.
(Note that $A_{\mu,2t}(2z,0) = \exp (-z^2/t)$ is never zero.)
Then we define
$$
  \mathcal{C}_{\mu,t}:= \{ f \in \mathcal{H} (\mathbb{C}^N) ~|~ Gf \in \mathcal{B}_{\mu,t/2} \},
$$
which becomes a Hilbert space with the inner product
$$
  \langle f_1, f_2 \rangle_{ \mathcal{C}_{\mu,t}} :=
   \langle Gf_1 , Gf_2 \rangle_{\mathcal{B}_{\mu,t/2}}
$$
for $f_1, f_2 \in \mathcal{C}_{\mu,t}$.
\end{definition}

The \textit{Version~C Hilbert space} $\mathcal{C}_{\mu,t}$ is a new generalized Segal-Bargmann space
in the context of the Dunkl theory associated to a finite Coxeter group.

\begin{theorem}
\label{version_c}
(Version C)
The transform $C_{\mu,t} : L^2(\mathbb{R}^N, \omega_{\mu,t}) \to \mathcal{C}_{\mu,t}$
is a unitary isomorphism.
Moreover, there are two relations between the kernel functions of Version~C and Version~A of the Segal-Bargmann transform.
One relation is given by
\begin{equation}
\label{C-A1}
        C_{\mu,t}(z,q) = A_{\mu,2t}(z,0) A_{\mu,t/2}(z/2,q).
\end{equation}
The second relation is
\begin{equation}
\label{C-A2}
C_{\mu,t}(z,q) = A_{\mu,t} (0,q) A_{\mu,t} (z,q).
\end{equation}
Both of this identities are for $z \in \mathbb{C}^N$ and $q \in \mathbb{R}^N$.

Finally, Version~C of the Segal-Bargmann transform can be written as Dunkl
convolution with a Dunkl heat kernel followed by analytic continuation.
Specifically, we have for $x \in \mathbb{R}^N$ that
\begin{equation}
\label{conv_and_cont}
C_{\mu,t} \phi (x) = \left( \sigma_{\mu,t} \ast_{\mu,t} \phi \right) (x),
\end{equation}
where $\sigma_{\mu,t}(q) := \rho_{\mu,t}(0,q) = e^{-q^2/2t}$ for $q \in \mathbb{R}^N$
is the one variable Dunkl heat kernel.
Thus $C_{\mu,t} \phi (z)$ is the analytic continuation to $z \in \mathbb{C}^N$
of the Dunkl convolution of the heat kernel $\sigma_{\mu,t}$ with
$\phi$ as a function of $x \in \mathbb{R}^N$.
\end{theorem}

\vskip 0.2cm \noindent
\textbf{Remarks:}
The identity (\ref{C-A1}) is a generalization to this context of
identity (A.18) proved in \cite{HA} and identity (2.16) proved in \cite{SBS}.
These last two identities were our motivation for considering (\ref{C-A1}) in the first place.
After having proved (\ref{C-A1}) we looked for other identities of the same sort,
and this is how we found (\ref{C-A2}).

Note that the kernel $A_{\mu,t}(z,q)$ is invariant under the three simultaneous dilations
\begin{equation}
\label{simul-dil}
z \mapsto az, \quad q \mapsto aq, \quad t \mapsto a^2 t
\end{equation}
for any $a > 0$.
Thus there is no such dilation so that the factor $A_{\mu,t/2}(z/2,q)$
in (\ref{C-A1}) becomes a function directly of the point
$(z,q)$ in $\mathbb{C}^N \times \mathbb{R}^N$ rather than some function of that point.

An advantage that (\ref{C-A2}) has over (\ref{C-A1}) is that it expresses the kernel $ C_{\mu,t}$
evaluated at $(z,q)$ in terms of expressions that depend only on $z$ and $q$
and not on their dilations.
Also the same ``time'' parameter is used in all the terms.
Of course, along with (\ref{C-A1}) and (\ref{C-A2}), we have all the identities that
result by dilating the expressions on their right sides as in (\ref{simul-dil}).
As far as we are aware, identity (\ref{C-A2}) is a new result.
Except for  (\ref{C-A2}), all of the results of this theorem reduce to
those of \cite{HA} when $\mu \equiv 0 $ and of \cite{SBS} when $N=1$, up to normalizations.
Both (\ref{C-A1}) and (\ref{C-A2}) imply that knowledge of the kernel for Version A
gives us all the information for writing down the kernel for Version C.
So, in this context, Version C is determined by Version A.

Finally, we would like to comment on why we write (\ref{C-A2}) as
$$
C_{\mu,t}(z,q) = A_{\mu,t} (0,q) A_{\mu,t} (z,q)
$$
instead of as
$$
C_{\mu,t}(z,q) = e^{-q^2/4t} A_{\mu,t} (z,q)
$$
even though the second formula is totally correct.
The point is that in the second formula there appears in an \textit{ad hoc}
manner the factor $e^{-q^2/4t}$, while in the first formula one sees
that this factor on the right comes from Version A, just as the second factor does.
So the first formula reveals more clearly the structure of the relation between
Versions A and C.
Similar comments apply to (\ref{C-A1}).

\vskip 0.2cm \noindent
{\bf Proof of Theorem \ref{version_c}:}
To show the identity (\ref{C-A1}), we note first that (\ref{def-kernel-a}) implies
$$
A_{\mu,2t}(z, 0) = \exp \left( -  z^2/2(2t) \right) = e^{-z^2/4t}.
$$
Then applying (\ref{def-kernel-a}) again we have
\begin{eqnarray*}
&& A_{\mu,2t}(z,0) A_{\mu,t/2}(z/2,q)
\\
&=& e^{-z^2/4t}
 \exp \left( -\dfrac{1}{2} \left( \dfrac{z}{2}  \right)^2 \dfrac{2}{t}
- \dfrac{q^2}{4} \dfrac{2}{t} \right)
E_\mu\left( \dfrac{z}{2} \left(\dfrac{2}{t}\right)^{1/2} \! \! \! \! \! \! \! \! , \, \, \,
q  \left(\dfrac{2}{t} \right)^{1/2} \right)
\\
&=& e^{-z^2/4t} \exp \left( -z^2/4t - q^2/2t \right)
E_\mu \left( \dfrac{z}{(2t)^{1/2}} , \dfrac{2^{1/2}q}{t^{1/2}} \right)
\\
&=& \exp \left( -z^2/2t - q^2/2t \right)
E_\mu \left( \dfrac{z}{t^{1/2}} , \dfrac{q}{t^{1/2}} \right)
\\
&=& \rho_{\mu,t} (z,q) = C_{\mu,t} (z,q)
\end{eqnarray*}
for all $z \in \mathbb{C}^N$ and $q \in \mathbb{R}^N$.

To show (\ref{C-A2}) we again use (\ref{def-kernel-a}) to calculate
\begin{eqnarray*}
A_{\mu,t} (0,q) A_{\mu,t} (z,q) &=& \left(
e^{-q^2/4t} \right) \left( e^{-z^2/2t - q^2/4t}
E_\mu \left( \dfrac{z}{t^{1/2}} , \dfrac{q}{t^{1/2}} \right) \right)
\\
&=& e^{-z^2/2t - q^2/2t}
E_\mu \left( \dfrac{z}{t^{1/2}} , \dfrac{q}{t^{1/2}} \right)
\\
&=& \rho_{\mu,t} (z,q)
= C_{\mu,t} (z,q).
\end{eqnarray*}

Next, we use definitions and (\ref{C-A1}) to evaluate for
$\psi \in  L^2(\mathbb{R}^N, \omega_{\mu,t})$ and $z \in \mathbb{C}^N$ that
\begin{eqnarray*}
(G C_{\mu,t} \psi ) (z) &=& 2^{\gamma_\mu/2 + N/4} \dfrac{ (C_{\mu,t} \psi)(2z) }{A_{\mu,2t}(2z,0)} \\
&=&  \dfrac{2^{\gamma_\mu/2 + N/4}}{A_{\mu,2t}(2z,0)} \int_{\mathbb{R}^N} \mathrm{d} \omega_{\mu,t}(q)
C_{\mu,t}(2z,q) \psi(q) \\
&=& \dfrac{2^{\gamma_\mu/2 + N/4}}{A_{\mu,2t}(2z,0)} \int_{\mathbb{R}^N} \mathrm{d} \omega_{\mu,t}(q)
A_{\mu,2t}(2z,0) A_{\mu,t/2} (z, q) \psi(q) \\
&=& 2^{\gamma_\mu/2 + N/4}
\int_{\mathbb{R}^N} \mathrm{d} \omega_{\mu,t/2}(q) 2^{-(\gamma_\mu + N/2)}
A_{\mu,t/2}(z,q) \psi(q) \\
&=& 2^{-(\gamma_\mu/2 + N/4)} \int_{\mathbb{R}^N} \mathrm{d} \omega_{\mu,t/2}(q) A_{\mu,t/2} (z,q) \psi(q) \\
&=& 2^{-(\gamma_\mu/2 + N/4)} A_{\mu,t/2} \psi(z).
\end{eqnarray*}
This may appear to be in error, since we have defined $A_{\mu,t/2}$ on the domain
$L^2(\mathbb{R}^N, \omega_{\mu,t/2})$ but we are evaluating it at
$\psi \in  L^2(\mathbb{R}^N, \omega_{\mu,t})$, since this is the domain of $C_{\mu,t}$.
However, the {\em set} of functions in $L^2(\mathbb{R}^N, \omega_{\mu,t})$ does not
depend on the value of $t>0$.
What changes when one varies $t$ is the measure, and consequently the inner product
and the norm, all of which only change by a positive real factor.
Explicitly, the relations for the measure and inner product for the values $s$ and $t$ are
$$
\mathrm{d}\omega_{\mu,t}(q) = \left( \dfrac{s}{t} \right)^{\gamma_\mu + N/2}
\mathrm{d}\omega_{\mu,s}(q)
$$
and
$$
\langle \psi, \phi \rangle_{L^2(\omega_{\mu,t})} = \left( \dfrac{s}{t} \right)^{\gamma_\mu + N/2}
\langle \psi, \phi \rangle_{L^2(\omega_{\mu,s})}.
$$
We note in passing that for fixed $N$ and variable $t$ the spaces
$\mathcal{B}_{\mu,t} $ considered as sets of holomorphic functions
on $\mathbb{C}^N$ are {\em not} equal in the case when $\mu \equiv 0$ nor when $N=1$.

In summary, the integral that defines $A_{\mu,t/2} \psi$ converges absolutely.
Then we have
\begin{eqnarray*}
\langle C_{\mu,t} \psi_1 ,  C_{\mu,t} \psi_2 \rangle_{\mathcal{C}_{\mu,t}} &=&
\langle GC_{\mu,t} \psi_1 ,  GC_{\mu,t} \psi_2 \rangle_{\mathcal{B}_{\mu,t/2}} \\
&=& \langle 2^{-(\gamma_\mu/2 + N/4)} A_{\mu,t/2} \psi_1 , 2^{-(\gamma_\mu/2 + N/4)}
A_{\mu,t/2} \psi_2 \rangle_{\mathcal{B}_{\mu,t/2}} \\
&=& 2^{-(\gamma_\mu + N/2)} \langle \psi_1 , \psi_2 \rangle_{ L^2(\omega_{\mu,t/2} ) }\\
&=& 2^{-(\gamma_\mu + N/2)}
\left( 2^{\gamma_\mu + N/2} \langle \psi_1 , \psi_2 \rangle_{ L^2(\omega_{\mu,t} ) } \right) \\
&=& \langle \psi_1 , \psi_2 \rangle_{ L^2(\omega_{\mu,t} ) }
\end{eqnarray*}
for $\psi_1, \psi_2 \in  L^2(\omega_{\mu,t} ) $,
and this shows that $C_{\mu,t}$ is a unitary transform from $ L^2(\mathbb{R}^N, \omega_{\mu,t})$
onto its range.

To show that $C_{\mu,t}$ is onto $\mathcal{C}_{\mu,t}$, we note that the range of
$C_{\mu,t}$ consists of all $f \in \mathcal{H} (\mathbb{C}^N)$ such that
$f = C_{\mu,t} \psi$ for some $\psi \in  L^2(\omega_{\mu,t} )$.
Then we note that we have these equivalences:
\begin{eqnarray*}
   f = C_{\mu,t} \psi &\Longleftrightarrow& Gf = GC_{\mu,t} \psi =
       2^{-(\gamma_\mu/2 + N/4)} A_{\mu,t/2} \psi \\
&\Longleftrightarrow& Gf \in \mathrm{Ran} (A_{\mu,t/2} ) \\
&\Longleftrightarrow& Gf \in \mathcal{B}_{\mu,t/2} \\
&\Longleftrightarrow& f \in \mathcal{C}_{\mu,t},
\end{eqnarray*}
which shows that the range of the transform $C_{\mu,t}$ is the space $\mathcal{C}_{\mu,t}$.

To show (\ref{conv_and_cont})
we note that
\begin{eqnarray}
\label{conv_and_cont_2}
C_{\mu,t} \phi (x) &=&
  \int_{ \mathbb{R}^N } \mathrm{d} \omega_{\mu,t}(q) C_{\mu,t} (x,q) \phi(q) \nonumber \\
  &=&
  \int_{ \mathbb{R}^N } \mathrm{d} \omega_{\mu,t}(q) \rho_{\mu,t} (x,q) \phi(q) \nonumber \\
  &=&
  \int_{ \mathbb{R}^N } \mathrm{d} \omega_{\mu,t}(q) \rho_{\mu,t} (q,x) \phi(q) \nonumber \\
 &=& \int_{ \mathbb{R}^N } \mathrm{d} \omega_{\mu,t}(q) \, \left( \mathcal{T}_{\mu,q}
\sigma_{\mu,t} \right) (x) \, \, \phi(q) \nonumber \\
 &=& \left( \sigma_{\mu,t} \ast_{\mu,t} \phi \right) (x).
\end{eqnarray}
So, it suffices to show for $q \in \mathbb{R}^N$
and $x \in \mathbb{R}^N$ that
\begin{equation}
\label{tsigma}
        \mathcal{T}_{\mu,x} \sigma_{\mu,t} (q) = \rho_{\mu,t} (x,q),
\end{equation}
since all the other steps in the chain of equalities in (\ref{conv_and_cont_2})
are immediate.

Actually, the identity (\ref{tsigma}) is known.
(See \cite{MR} or \cite{TX} for instance.)
However, for completeness we prove it.
We begin with a more well known identity for the Dunkl kernel,
$$
 \int_{\mathbb{R}^N} \mathrm{d} \omega_{\mu,1} (x) \, ( e^{-\Delta_\mu/2} p ) (x) \,
 E_\mu(x,w) \, e^{-x^2/2} = e^{w^2/2} p(w),
$$
which is equation (2.4) from \cite{SBSBO} in our notation.
(Recall that the Macdonald-Mehta-Selberg constant in \cite{SBSBO} has been
absorbed into the measure.)
Here, $\mu \ge 0$, $w \in \mathbb{C}^N $ and $p(x)$ is a polynomial in $x = (x_1, \dots, x_N)$.
In particular, we take $p(x) \equiv 1$, a constant polynomial, and $w=-ik$
for $k \in \mathbb{R}^N $.
So we get
$$
\int_{\mathbb{R}^N} \mathrm{d} \omega_{\mu,1} (x) \,
 E_\mu(x,-ik) \, e^{-x^2/2} = e^{-k^2/2}.
$$
Now, by dilating $x \mapsto x/t^{1/2}$ and $k \mapsto k/t^{1/2}$
this becomes
$$
\int_{\mathbb{R}^N} \mathrm{d} \omega_{\mu,t} (x) \,
 E_\mu \left(\dfrac{x}{t^{1/2}},\dfrac{-ik}{t^{1/2}} \right)  e^{-x^2/2t}
 = e^{-k^2/2t}
$$
or equivalently
$$
  \mathcal{F}_{\mu,t} \sigma_{\mu,t} (k) = e^{-k^2/2t} =  \sigma_{\mu,t} (k).
$$
Recalling that (\ref{def_mu_trans}) says that
$$
\mathcal{T}_{\mu,x} =
\mathcal{F}_{\mu,t}^* E_\mu \left( -x, \dfrac{ik}{t} \right) \mathcal{F}_{\mu,t},
$$
we have that
$$
\mathcal{T}_{\mu,x} \sigma_{\mu,t} (q) =
\int_{\mathbb{R}^N} \mathrm{d} \omega_{\mu,t} (k) \,
 E_\mu \left( \dfrac{k}{t^{1/2}}, \dfrac{iq}{t^{1/2}}   \right)
 E_\mu \left( -\dfrac{ix}{t^{1/2}}, \dfrac{k}{t^{1/2}} \right) e^{-k^2/2t}.
$$
Now equation (2.5) in \cite{SBSBO} in our notation is
$$
\int_{\mathbb{R}^N} \mathrm{d} \omega_{\mu,1} (k) \, E_\mu (k,z) E_\mu (k,w) e^{-k^2/2}
=  e^{(z^2 + w^2)/2} E_\mu (z,w)
$$
for $z,w \in \mathbb{C}^N$ and $\mu \ge 0$.
Substituting $z=iq/t^{1/2}$ and $w = -i x/t^{1/2}$ and dilating the variable of integration
by $k \mapsto k/t^{1/2}$ in the last equation gives us
\begin{eqnarray*}
&& \int_{\mathbb{R}^N} \mathrm{d} \omega_{\mu,t} (k) \,
 E_\mu \left( \dfrac{k}{t^{1/2}}, \dfrac{iq}{t^{1/2}}   \right)
 E_\mu \left( \dfrac{k}{t^{1/2}} , -\dfrac{ix}{t^{1/2}} \right) e^{-k^2/2t} \\
&=& e^{-(q^2 + x^2)/2t} E_\mu \left( \dfrac{q}{t^{1/2}} , \dfrac{x}{t^{1/2}} \right).
\end{eqnarray*}
Putting this together with the equation above and
using the symmetry of the Dunkl kernel $E_\mu$ yields
$$
\mathcal{T}_{\mu,x} \sigma_{\mu,t} (q) =
 e^{-(x^2 + q^2)/2t} E_\mu \left( \dfrac{x}{t^{1/2}} , \dfrac{q}{t^{1/2}} \right)
 = \rho_{\mu,t}(x,q)
$$
for $x,q \in \mathbb{R}^N$.
This finishes the proof of (\ref{tsigma}), which
is the last assertion to be proved. $\quad \blacksquare$

We feel it is rather important to comment on the definitions
(\ref{def_B_mu_t}) of the transform $B_{\mu, t} $ and
(\ref{def_C_mu_t}) of the transform $C_{\mu, t} $.
We note that
$$
B_{\mu,t} \phi (z) =
  \int_{ \mathbb{R}^N } \mathrm{d} m_{\mu,t}(q) B_{\mu,t} (z,q) \phi(q) =
 \int_{ \mathbb{R}^N } \mathrm{d} \omega_{\mu,t}(q) \rho_{\mu,t} (z,q) \phi(q)
$$
by using the definitions of $\mathrm{d} m_{\mu,t}(q)$ and $ B_{\mu,t}(z,q)$.
Also by definitions we have that
$$
C_{\mu,t} \phi (z) =
  \int_{ \mathbb{R}^N } \mathrm{d} \omega_{\mu,t}(q) \rho_{\mu,t} (z,q) \phi(q)
$$
While these last two formulas seem to be equal, this is extremely misleading.
They are really quite different formulas.
The point is that the domains of the transforms $B_{\mu, t} $ and $C_{\mu, t} $ are not
equal, but are different Hilbert spaces.
Similarly, the ranges of $B_{\mu, t} $ and $C_{\mu, t} $ are not equal,
 but are different Hilbert spaces.
Since each one of the transforms $B_{\mu, t} $ and $C_{\mu, t} $ is a unitary transform from
its domain to its range, these domains and ranges (as Hilbert spaces) are
of central importance to the structure of the theory.
However, it is noteworthy that there is this formal relation
(that one might call ``symbolically identical'') between
(\ref{def_B_mu_t}) and (\ref{def_C_mu_t}), which is completely analogous to the formal relation
between Versions B and C in \cite{HA}.
Moreover, we call again to the reader's attention that in our convention
the integral kernel functions for the transforms
$B_{\mu, t}$ and $C_{\mu, t}$ are \textit{not} equal.

In this same vein the Segal-Bargmann space and corresponding Segal-Bargmann transform
in \cite{SBSBO} and \cite{SO} correspond to our Version~A, and not to Version~C, even though
Version~C is completely determined by Version~A, as we noted earlier.
The point here is that neither the range space of $C_{\mu, t} $ nor the transform  $C_{\mu, t} $
itself (see (\ref{def_C_mu_t})) really appears
in \cite{SBSBO} or \cite{SO}, except with an unreasonably highly perceptive hindsight.
Of course, the domain space of (\ref{def_C_mu_t}) does appear in \cite{SBSBO} and \cite{SO},
but we think this is far from adequate for supporting the assertion that Version C
is presented in \cite{SBSBO} or \cite{SO}.
We think that Version~C (as well as Version~B) in the Coxeter context
was first introduced by us in \cite{SBS}, though only in the one-dimensional case.

\section{Conclusion}
It appears to us that the results of this article depend on the Euclidean space
structure of the configuration space $\mathbb{R}^N$ and
the phase space $\mathbb{C}^N$ in great part through the dilations defined on
these vector spaces.
The configuration space also carries a one-parameter family of measures, namely
the measures $\mathrm{d} \omega_{\mu,t}$ for $t>0 $, and the dilations are
related in a very simple way to these measures.

The problem of finding a measure on the phase space $\mathbb{C}^N$
(to realize the Segal-Bargmann space as the holomorphic $L^2$ space
for that measure) is complicated
by the fact that when $N=1$ and $\mu \ne 0$ it is known that the Segal-Bargmann
space for a given $t > 0$
is realized by using {\em two} measures on the phase space or,
equivalently, as a closed subspace of
the space of $L^2$ holomorphic functions on $\mathbb{C} \times \{ -1, 1 \} $
for some measure on it. (See \cite{SBS2}.)
Under some rather restrictive hypotheses, Asai has shown in \cite{AS} for dimension $N=1$
that the Segal-Bargmann space associated to a probability measure
on the configuration space $\mathbb{R}$
can be realized as the $L^2$ space of holomorphic functions on the phase space $\mathbb{C}$
for a unique probability measure on $\mathbb{C}$.
The case $N=1$ (with $G = \mathbb{Z}_2$ and $\mu \ne 0$) of Version B of the theory presented here
is not included among the cases considered in \cite{AS}.
Moreover, we thank L.~Echavarr\'ia \cite{EC} for showing us recently that for $N=1$ and $\mu \ne 0$
there is no way to realize version A (which is the same as Version B) of the Segal-Bargmann space
associated with the Coxeter group $\mathbb{Z}_2$ as the holomorphic functions in
$L^2 ( \mathbb{C}, \nu \,\mathrm{d} x \, \mathrm{d} y)$ for a positive density
function $\nu : \mathbb{C} \to (0,\infty)$.

So a very general open problem is to identify exactly when the codomain Hilbert space
of a Segal-Bargmann type transform can be realized
as a space of holomorphic $L^2$ functions on the phase space
with respect to some measure on the phase space or if a representation using
a set of measures can be found.
And a subsequent problem when such a measure or measures exist is that of their uniqueness.
For example, as far as we know, it might be possible to represent the Segal-Bargmann space
for $N=1$ and $\mu \ne 0$ with three or more measures in an essentially new way.

A curious point in the Dunkl theory is that the one variable Dunkl heat kernel
$\sigma_{\mu, t}(x) = e^{-x^2/2t}$ for $x \in \mathbb{R}^N$
does not depend on $\mu$ and so is identical (given our conventions and normalizations)
with the classical one variable heat kernel for the usual Laplacian in Euclidean space.
Of course, the two variable Dunkl heat kernel $\rho_{\mu, t}$ \textit{does} depend on $\mu$, since
it is obtained from the one variable Dunkl heat kernel by a $\mu$ dependent Dunkl translation.
(See (\ref{tsigma}).)
We wonder whether more might be said about this.

Finally, let us note that our results have been proved when $\mu \ge 0$.
It might be possible to weaken this hypothesis while still having the same results.

\section{Acknowledgments}
This article was begun while I was visiting the University of Virginia.
Thanks go to all those who made that a pleasant and productive experience,
but more than anybody I wish to thank my host there, Larry Thomas.
I also thank N.~Asai for telling me about his article \cite{AS},
M.~Castillo Salgado for bringing reference \cite{TX} to my attention,
L.~Echavarr\'ia for allowing me to mention his result in the Conclusion
and B.~Hall for valuable comments.

\end{document}